\documentclass[10pt]{iopart}

\usepackage{cite} 

\begin{document}

\newcommand{\dd}{{\rm{d}}}       
\newcommand{\erovno}{& \equiv &} 
\newcommand{\rovno}{& = &}       
\newcommand{\ssqrt}{{\textstyle\frac1{\sqrt{2}}}} 
\newcommand{\df}{{\phi}}         

\newcommand{\boldu}{\mbox{\boldmath$u$}} 
\newcommand{\bolde}{\mbox{\boldmath$e$}} 
\newcommand{\boldk}{\mbox{\boldmath$k$}} 
\newcommand{\boldl}{\mbox{\boldmath$l$}} 
\newcommand{\boldm}{\mbox{\boldmath$m$}} 
\newcommand{\bboldm}{\bar{\mbox{\boldmath$m$}}} 
\newcommand{\boldZ}{\mbox{\boldmath$Z$}} 
\newcommand{\boldT}{\mbox{\boldmath$T$}} 
\newcommand{\boldF}{\mbox{\boldmath$F$}} 

\title[Algebraic structure of Robinson--Trautman and Kundt geometries]{Algebraic structure of Robinson--Trautman and Kundt geometries in arbitrary dimension}

\author{J Podolsk\'y and R \v{S}varc}

\address{Institute of Theoretical Physics, Faculty of Mathematics and Physics, Charles University in Prague, V~Hole\v{s}ovi\v{c}k\'ach~2, 180~00 Praha 8, Czech Republic }
\eads{\mailto{podolsky@mbox.troja.mff.cuni.cz} and \mailto{robert.svarc@mff.cuni.cz}}

\begin{abstract}
We investigate the Weyl tensor algebraic structure of a fully general family of $D$-dimensional geometries that admit a non-twisting and shear-free null vector field~${\mathbf{k}}$. From the coordinate components of the curvature tensor we explicitly derive all Weyl scalars of various boost weights. This enables us to give a complete algebraic classification of the metrics in the case when the optically privileged null direction~${\mathbf{k}}$ is a (multiple) Weyl aligned null direction (WAND). No field equations are applied, so that the results are valid not only in Einstein's gravity, including its extension to higher dimensions, but also in any metric gravitation theory that admits non-twisting and shear-free spacetimes.

We prove that all such geometries are of type~I(b), or more special, and we derive surprisingly simple necessary and sufficient conditions under which ${\mathbf{k}}$ is a double, triple or quadruple WAND. All possible algebraically special types, including the refinement to subtypes, are thus identified, namely II(a), II(b), II(c), II(d), III(a), III(b), N, O, II$_i$, III$_i$, D, D(a), D(b), D(c), D(d), and their combinations. Some conditions are identically satisfied in four dimensions.

We discuss both important subclasses, namely the Kundt family of geometries with the vanishing expansion (${\Theta=0}$) and the Robinson--Trautman family (${\Theta\not=0}$, and in particular ${\Theta=1/r}$). Finally, we apply Einstein's field equations and obtain a classification of all Robinson--Trautman vacuum spacetimes. This reveals fundamental algebraic differences in the ${D>4}$ and ${D=4}$ cases, namely that in higher dimensions there only exist such spacetimes of types D(a)$\equiv$D(abd), D(c)$\equiv$D(bcd) and O.
\end{abstract}

\submitto{\CQG}


\maketitle

\section{Introduction}
\label{intro}

Exact spacetimes play a crucial role in understanding Einstein's general relativity and other metric theories of gravity. They enable us to investigate many mathematical and physical aspects of fundamental models in cosmology, black hole physics and theory of gravitational waves.

Among the most important of such classes of exact solutions there are Robinson--Trautman \cite{RobTra60,RobTra62} and Kundt \cite{Kundt:1961,Kundt:1962} geometries. They were discovered almost simulta\-neously about half a century ago, shortly after the advent of new concepts and techniques in general relativity, in particular geometrical optics of null congruences and algebraic classification of the Weyl tensor. Since then, an enormous progress has been made in investigation of their various properties.

From the geometrical point of view, both these classes belong to a large family of geometries admitting a non-twisting shear-free congruence of geodesics, generated by a null vector field ${\mathbf{k}}$. The Kundt class is defined by having vanishing expansion while the other case with non-vanishing expansion defines the Robinson--Trautman class. The former includes the famous pp-waves (with a covariantly constant ${\mathbf{k}}$), more general non-expanding gravitational waves (including gyratons, non-vanishing cosmological constant $\Lambda$, impulsive limits), VSI and CSI spacetimes (for which all scalar invariants of curvature vanish and are constant, respectively), or the direct-product spacetimes (Bertotti--Robinson, Nariai, Pleba\'{n}ski--Hacyan). In the Robinson--Trautman class there are, e.g., some well-known black holes (Schwarzchild, Reissner--Nordstr\"om, Schwarzchild--de Sitter, Vaidya), expanding spherical gravitational waves (including $\Lambda$), the C-metric (representing the field of accelerated black holes) or Kinnersley's and Bonnor's ``photon rockets''. Details and a number of references can be found in the monographs \cite{Stephani:2003,GriffithsPodolsky:2009} (chapters 28, 31 and 19, 18, respectively).

In view of the growing interest to generalize Einstein's theory and to extend it to higher dimensions, it is a natural task to find and analyse specific properties of such spacetimes. Assuming the validity of Einstein's field equations (for vacuum with $\Lambda$, aligned electromagnetic field, pure radiation, gyratonic matter), the explicit Robinson--Trautman class in any dimension was studied in \cite{PodOrt06,OrtPodZof08,SvarcPodolsky:2014}. The complementary Kundt class was also investigated, e.g., in \cite{PodolskyZofka:2009,ColeyEtal:2009}. The results were summarized in the recent review  \cite{OrtaggioPravdaPravdova:2013} on algebraic properties of higher dimensional spacetimes.

In this paper we consider the fully general class of non-twisting and shear-free geometries in an arbitrary dimension ${D\ge4}$, without assuming any field equations. Starting from the canonical form of the metric we derive all components of the Riemann, Ricci and Weyl tensors. These are projected onto a suitable null frame. It enables us to give a complete and explicit classification of the whole class into the algebraic types and subtypes based on the WAND multiplicity of the optically privileged null vector field ${\mathbf{k}}$. Our new results thus considerably generalize the study of algebraic structure of the non-expanding Kundt family of geometries \cite{PodolskySvarc:2013a}.

We introduce the general metric in section~\ref{sec_geom}. In section~\ref{genralPsi} the null frame components of the Weyl tensor are  employed for the algebraic classification.  The necessary and sufficient conditions for all principal and secondary alignment (sub)types are discussed in sections~\ref{multipleWAND} and~\ref{otherWANDl}. Results for the Kundt class are summarized in section~\ref{Kundt_geometries} while those for the Robinson--Trautman class are contained in section~\ref{RT_geometries}. In section~\ref{RTexample} we discuss a special case, namely the Robinson--Trautman vacuum spacetimes in $D$-dim Einstein's theory. Coordinate components of the Riemann, Ricci, and Weyl tensors for the generic non-twisting shear-free geometry are given~in~\ref{appendixA}.

\section{Robinson--Trautman and Kundt geometries}
\label{sec_geom}

The metric of the most general non-twisting $D$-dimensional geometry can be written in the form \cite{PodOrt06}
\begin{equation}
\fl \dd s^2 = g_{pq}(r,u,x)\, \dd x^p\,\dd x^q+2\,g_{up}(r,u,x)\, \dd u\, \dd x^p -2\,\dd u\,\dd r+g_{uu}(r,u,x)\, \dd u^2 \,, \label{general nontwist}
\end{equation}
if natural coordinates are used. A non-twisting character of the spacetime implies the existence of a foliation by null hypersurfaces ${u=\,}$const., i.e., a family of maximal integral submanifolds labeled by the coordinate $u$. By the Frobenius theorem, this is equivalent to the existence of a non-twisting null vector field ${\mathbf{k}}$ that is everywhere tangent (and normal) to ${u=\,}$const. Since this field ${\mathbf{k}}$ generates a congruence of null geodesics in the whole spacetime, it is most natural to take their affine parameter~$r$ as the second coordinate, so that ${\mathbf{k}=\mathbf{\partial}_r}$. At any fixed $u$ and $r$ we are thus left with a ${(D-2)}$-dimensional Riemannian manifold covered by the spatial coordinates ${x^p}$. We will use the indices $m,n,p,q$ (ranging from $2$ to ${D-1}$) to label these spatial coordinates on the transverse space, and a shorthand $x$ for their complete set. The nonvanishing contravariant metric components are \begin{equation}
g^{pq}\,, \quad g^{ru}=-1\,, \quad g^{rp}= g^{pq}g_{uq} \,,
\quad g^{rr}= -g_{uu}+g^{pq}g_{up}g_{uq} \,, \label{ContravariantMetricComp}
\end{equation}
where $g^{pq}$ is an inverse matrix to $g_{pq}$. This implies
\begin{equation}
g_{up}= g_{pq}g^{rq} \,, \quad g_{uu}= -g^{rr}+g_{pq}g^{rp}g^{rq} \,. \label{CovariantMetricComp}
\end{equation}

The covariant derivative of the geometrically privileged null vector field ${\mathbf{k}=\mathbf{\partial}_r}$ with respect to the metric~(\ref{general nontwist}) is ${k_{a;b}=\Gamma^{u}_{ab}=\frac{1}{2}g_{ab,r}}$ so that ${k_{r;b}=0=k_{a;r}}$. Consequently, the \emph{optical matrix}~\cite{OrtaggioPravdaPravdova:2013} defined as ${\rho_{ij}\equiv k_{a;b}\,m_i^am_j^b}$, where ${m_i^a}$ are components of ${(D-2)}$ unit vectors
$\boldm_i$ such that  ${\boldm_i\cdot\mathbf{k}=0\Rightarrow m_i^u=0}$, forming an orthonormal basis in the transverse Riemannian space, is simply given by ${\rho_{ij}=k_{p;q}\,m^p_im^q_j= \frac{1}{2} g_{pq,r}\,m^p_im^q_j}$. This can be decomposed as ${\rho_{ij}=A_{ij}+\sigma_{ij}+\Theta\delta_{ij}}$, where ${A_{ij}\equiv\rho_{[ij]}}$ is the antisymmetric \emph{twist matrix}, ${\sigma_{ij}}$ is the symmetric traceless \emph{shear matrix}, and the scalar ${\Theta\equiv\frac{1}{D-2}\,\delta^{ij}\rho_{ij}}$ determines the \emph{expansion} of the privileged vector field ${\mathbf{k}}$.

It can be observed that ${A_{ij}=\frac{1}{2} g_{pq,r}\,m^p_{[i}m^q_{j]}=0}$ which confirms that the metric~(\ref{general nontwist}) is \emph{non-twisting}. Now, imposing the additional condition that the metric is \emph{shear-free}, ${\sigma_{ij}=0}$, we obtain the relation
\begin{equation}
\rho_{ij}=\Theta\delta_{ij}={\textstyle\frac{1}{2}}\,g_{pq,r}m^p_im^q_j \,. \label{OptMat}
\end{equation}
Using the orthonormality relation ${\delta_{ij}=g_{pq}m_i^pm_j^q}$ we thus immediately infer
\begin{equation}
g_{pq,r}=2\Theta g_{pq} \,,
\label{shearfree condition}
\end{equation}
implying ${g_{pq,rr}=2\big(\Theta_{,r}+2\Theta^2\big)g_{pq}}$. The expression (\ref{shearfree condition}) can be integrated as
\begin{equation}
g_{pq}=R^2(r,u,x)\,h_{pq}(u,x) \,,
\qquad \hbox{where}\quad \frac{R_{,r}}{R}=\Theta \,. \label{IntShearFreeCond}
\end{equation}
Since either ${\Theta = 0}$ or ${\Theta \neq 0}$, there are thus two distinct classes of non-twisting shear-free geometries. The \emph{Kundt class} \cite{Kundt:1961,Kundt:1962,Stephani:2003,GriffithsPodolsky:2009,PodolskyZofka:2009,OrtaggioPravdaPravdova:2013,PodolskySvarc:2013a,PodolskySvarc:2013b} is defined by having the vanishing expansion, ${\Theta=0}$, in which case the spatial metric ${g_{pq}(u,x)=h_{pq}(u,x)}$ is independent of the affine parameter $r$ (and $R$ in (\ref{IntShearFreeCond}) effectively reduces to ${R=1}$). The other case ${\Theta\neq 0}$ gives the expanding \emph{Robinson--Trautman class} \cite{RobTra60,RobTra62,Stephani:2003,GriffithsPodolsky:2009,PodOrt06,OrtPodZof08,OrtaggioPravdaPravdova:2013}, for which $R$ is a non-trivial function of $r$ determined by
${R=\exp\big(\int\Theta(r,u,x)\,\dd r\big)}$.

\section{Frame components of the Weyl tensor and its classification}
\label{genralPsi}

The natural null frame for the general metric (\ref{general nontwist}) is given by
\begin{equation}
{\textstyle \boldk=\mathbf{k}=\mathbf{\partial}_r \, , \ \quad \boldl=\frac{1}{2}g_{uu}\mathbf{\partial}_r+\mathbf{\partial}_u \, , \ \quad \boldm_i=m_i^p(g_{up}{\partial}_r+\mathbf{\partial}_p)} \,, \label{nat null frame}
\end{equation}
satisfying ${\boldk\cdot\boldl=-1}$, ${\boldm_i\cdot\boldm_j=\delta_{ij}}$.
All the Weyl tensor components with respect to such a frame ${(\boldk,\,\boldl,\,\boldm_i)}$, sorted by the boost weight, can be denoted as \cite{PodolskySvarc:2012,PodolskySvarc:2013a}
\begin{eqnarray}
\Psi_{0^{ij}} \rovno C_{abcd}\; k^a\, m_i^b\, k^c\, m_j^d \,, \nonumber \\
\Psi_{1^{ijk}}\rovno C_{abcd}\; k^a\, m_i^b\, m_j^c\, m_k^d \,,
\hspace{11mm} \Psi_{1T^{i}} = C_{abcd}\; k^a\, l^b\, k^c\, m_i^d \,\nonumber \\
\Psi_{2^{ijkl}} \rovno C_{abcd}\; m_i^a\, m_j^b\, m_k^c\, m_l^d \,,
\hspace{11mm} \Psi_{2S} = C_{abcd}\; k^a\, l^b\, l^c\, k^d \,,\nonumber \\
\Psi_{2^{ij}} \rovno C_{abcd}\; k^a\, l^b\, m_i^c\, m_j^d \,,
\hspace{12.0mm} \Psi_{2T^{ij}} = C_{abcd}\; k^a\, m_i^b\, l^c\, m_j^d \,, \nonumber \\
\Psi_{3^{ijk}} \rovno C_{abcd}\; l^a\, m_i^b\, m_j^c\, m_k^d \,,
\hspace{11.9mm} \Psi_{3T^{i}} = C_{abcd}\; l^a\, k^b\, l^c\, m_i^d \,,\nonumber\\
\Psi_{4^{ij}} \rovno C_{abcd}\; l^a\, m_i^b\, l^c\, m_j^d \,. \label{defPsiCoef}
\end{eqnarray}
The scalars in the right column could, in fact, be obtained from those in the left column by contractions, namely
${\Psi_{1T^i}=\Psi_{1^{k}}{}^{_k}{}_{^i}}$,
${\Psi_{2S}=\Psi_{2T^{k}}{}^{_k}}$,
${\Psi_{2T^{(ij)}}=\frac{1}{2}\Psi_{2^{ikj}}{}^{_k}}$,
${\Psi_{2T^{[ij]}}=\frac{1}{2}\Psi_{2^{ij}}}$,
${\Psi_{3T^i}=\Psi_{3^{k}}{}^{_k}{}_{^i}}$. Relations of these Newman--Penrose-like quantities to other equivalent notations employed in \cite{OrtaggioPravdaPravdova:2013} and elsewhere can be found in \cite{PodolskySvarc:2013a}.

For the invariant (sub)classification of the Weyl tensor algebraic structure it is also necessary to introduce the following irreducible components of these scalars (see \cite{OrtaggioPravdaPravdova:2013}):
\begin{eqnarray}
\fl \tilde{\Psi}_{1^{ijk}}\erovno{\textstyle
\Psi_{1^{ijk}}-\frac{1}{D-3}\big(\delta_{ij}\Psi_{1T^{k}}-\delta_{ik}\Psi_{1T^{j}}\big)} \,, \\
\fl \tilde{\Psi}_{2T^{(ij)}}\erovno{\textstyle \Psi_{2T^{(ij)}}-\frac{1}{D-2}\delta_{ij}\Psi_{2S}} \,, \\
\fl \tilde{\Psi}_{2^{ijkl}} \erovno{\textstyle \Psi_{2^{ijkl}}-\frac{2}{D-4}\big(\delta_{ik}\tilde{\Psi}_{2T^{(jl)}}+\delta_{jl}\tilde{\Psi}_{2T^{(ik)}}
-\delta_{il}\tilde{\Psi}_{2T^{(jk)}}-\delta_{jk}\tilde{\Psi}_{2T^{(il)}}\big)} \nonumber \\
\fl &&{\textstyle -\frac{2}{(D-2)(D-3)}\big(\delta_{ik}\delta_{jl}-\delta_{il}\delta_{jk}\big)\Psi_{2S}} \,, \\
\fl \tilde{\Psi}_{3^{ijk}}\erovno{\textstyle
\Psi_{3^{ijk}}-\frac{1}{D-3}\big(\delta_{ij}\Psi_{3T^{k}}-\delta_{ik}\Psi_{3T^{j}}\big)} \,.
\end{eqnarray}

The main step now is to project the coordinate components (\ref{Weyl rprq})--(\ref{Weyl upuq}) of the Weyl tensor of a generic non-twisting shear-free geometry (see~\ref{appendixA}) onto the null frame~(\ref{nat null frame}). A long calculation with non-trivial cancelations of various terms reveals that the corresponding Weyl scalars take the following explicit and surprisingly simple form
\begin{eqnarray}
\Psi_{0^{ij}} &=& 0 \,, \label{Psi0ij}\\
\Psi_{1T^{i}} &=& {\textstyle m_i^p\,\frac{D-3}{D-2}\,\big[(-\frac{1}{2}g_{up,r}+\Theta g_{up})_{,r}+\Theta_{,p}\big]} \,, \label{Psi1Tj}\\
\tilde{\Psi}_{1^{ijk}} &=& 0 \,, \label{Psi1ijk}\\
\Psi_{2S}&=& {\textstyle \frac{D-3}{D-1}\, P}\,, \label{Psi2s} \\
\tilde{\Psi}_{2T^{(ij)}}&=& {\textstyle m_i^pm_j^q\,\frac{1}{D-2}\,\big(Q_{pq}-\frac{1}{D-2}\,g_{pq}\,Q \big)} \,,\label{Psi2Tij} \\
\tilde{\Psi}_{2^{ijkl}} &=& m_i^m m_j^p m_k^n m_l^q\,\,^{S}C_{mpnq} \,, \label{Psi2ijkl}\\
\Psi_{2^{ij}} &=& m_i^pm_j^q\, F_{pq} \,, \label{Psi2ij}\\
\Psi_{3T^{i}} &=& {\textstyle m_i^p\,\frac{D-3}{D-2}\, V_p} \,, \label{Psi3Tj}\\
\tilde{\Psi}_{3^{ijk}} &=& {\textstyle m_i^pm_j^mm_k^q\,\big(X_{pmq}-\frac{2}{D-3}\,g_{p[m}X_{q]}\big)} \,, \label{Psi3ijk}\\
\Psi_{4^{ij}} &=& {\textstyle m_i^pm_j^q\,\big(W_{pq}-\frac{1}{D-2}\,g_{pq} W\big)} \,, \label{Psi4ij}
\end{eqnarray}
where
\begin{eqnarray}
\fl P \rovno {\textstyle \big(\frac{1}{2}g_{uu,r}-\Theta g_{uu}\big)_{,r}+\frac{1}{(D-2)(D-3)}\,^{S}R-\frac{1}{4}\frac{D-4}{D-2}\,g^{mn}g_{um,r}g_{un,r}} \nonumber \\
\fl && \hspace{0.0mm} {\textstyle +\frac{1}{D-2}\big(g^{rn}g_{un,rr}+g^{mn}g_{um,r||n}\big)-\frac{2}{D-2}\,g^{rn}g_{un}\Theta_{,r}-2\Theta_{,u}-\frac{4}{D-2}\,g^{rn}\Theta_{,n}} \nonumber \\
\fl && \hspace{0.0mm} {\textstyle -\Theta^2\frac{D-4}{D-2}\,g^{rn}g_{un}+\Theta\big(\frac{D-6}{D-2}\,g^{rn}g_{un,r}-\frac{2}{D-2}\,g^{mn}g_{um||n}\big)} \,, \label{P}\\
\fl Q_{pq} \rovno {\textstyle
\,^{S}R_{pq} +(D-4)\big[\frac{1}{2}\big(f_{pq}+g_{u(p}g_{q)u,rr}\big)
-\big(\Theta_{,r}-\Theta^2\big)g_{up}g_{uq}} \nonumber \\
\fl &&\hspace{26.0mm} {\textstyle -2g_{u(p}\Theta_{,q)}-\Theta\big(g_{u(p||q)}+2g_{u(p}g_{q)u,r}\big)\big]} , \label{Qpq}\\
\fl F_{pq} \rovno {\textstyle
g_{u[p,q],r}-g_{u[p}g_{q]u,rr}+2\Theta(g_{u[p}g_{q]u,r}-g_{u[p,q]})} \,, \label{Fpq}\\
\fl V_p \rovno{\textstyle
\frac{1}{2}\big[\frac{1}{2}g_{uu}g_{up,rr}-g_{uu,rp}+g_{up,ru}-\frac{1}{2}g^{rn}g_{un,r}g_{up,r}} \nonumber \\
\fl && \hspace{3.5mm} {\textstyle +g^{mn}g_{um,r}E_{np}-g_{up}\big(g_{uu,rr}-\frac{1}{2}g^{mn}g_{um,r}g_{un,r}\big)\big]} \nonumber \\
\fl && \hspace{0.0mm} {\textstyle +\frac{1}{D-3}\big[\frac{1}{2}g^{rn}g_{un}g_{up,rr}+g^{mn}e_{m[n}g_{p]u,r}-g^{rn}g_{u[n,p],r}+\frac{1}{2}g^{rn}\big(g_{u[p,r||n]}+f_{pn}\big)} \nonumber \\
\fl && \hspace{11.0mm} {\textstyle -g^{mn}\big(g_{m[p,u||n]}+g_{u[m,p]||n}\big)-\frac{1}{2}g_{up}\big(g^{rn}g_{un,rr}+g^{mn}f_{mn}\big)\big]} \nonumber \\
\fl && \hspace{0.0mm} {\textstyle +\frac{1}{2}g_{up}g_{uu}\Theta_{,r}+g_{up}\Theta_{,u}+\frac{1}{2}g_{uu}\Theta_{,p}} \nonumber \\
\fl && \hspace{0.0mm} {\textstyle
-\Theta \big[\,\frac{1}{2}g_{uu}g_{up,r}-g_{uu,p}+g_{up,u}
-g^{rn}g_{u[n}g_{p]u,r}+g^{rn}E_{np}-g_{up}g_{uu,r}} \nonumber \\
\fl && \hspace{7.0mm} {\textstyle +\frac{1}{D-3}(3g^{rn}g_{u[n}g_{p]u,r}-3g^{rn}g_{u[n,p]}-\frac{1}{2}g_{up}g^{mn}g_{mn,u}+\frac{1}{2}g^{rn}g_{np,u})\big]} \,, \label{Vp}\\
\fl X_{pmq} \rovno {\textstyle g_{p[m,u||q]}+g_{u[q,m]||p}+g_{up}g_{u[m}g_{q]u,rr}+e_{p[m}g_{q]u,r}} \nonumber \\
\fl && \hspace{0.0mm} {\textstyle -g_{u[q}g_{m]u,r||p}-g_{up}g_{u[m,r||q]}-\frac{1}{2}g_{u[q}g_{m]u,r}g_{up,r}} \nonumber \\
\fl && \hspace{0.0mm} {\textstyle +\Theta\big(3g_{u[q}g_{m]u,r}g_{up}+g_{u[q}g_{m]p,u}+g_{u[q}g_{m]u||p}-g_{up||[m}g_{q]u}-2g_{u[q,m]}g_{up}\big)} \,, \label{Xpmq}\\
\fl W_{pq} \rovno {\textstyle -\frac{1}{2}g_{uu||p||q}-\frac{1}{2}g_{pq,uu}+g_{u(p,u||q)}-\frac{1}{2}g_{uu,r}e_{pq}
+\frac{1}{2}g_{uu,(p}g_{q)u,r}-g_{uu,r(p}g_{q)u}} \nonumber \\
\fl && \hspace{0.0mm} {\textstyle
+\frac{1}{2}g_{uu}g_{u(p,r||q)}+\frac{1}{2}g_{uu}g_{u(q}g_{p)u,rr}
-\frac{1}{2}g_{uu,rr}g_{up}g_{uq}+g_{u(q}g_{p)u,ru} } \nonumber \\
\fl && \hspace{0.0mm} {\textstyle +\frac{1}{4}g^{mn}\big(g_{um}g_{un}g_{up,r}g_{uq,r}+g_{um,r}g_{un,r}g_{up}g_{uq}\big)
-\frac{1}{2}g^{mn}g_{um}g_{un,r}g_{u(q}g_{p)u,r}} \nonumber \\
\fl && \hspace{0.0mm} {\textstyle +g^{mn}\big(E_{mp}E_{nq}+g_{um,r}E_{n(p}g_{q)u}-g_{um}E_{n(p}g_{q)u,r}\big)} \nonumber \\
\fl && \hspace{0.0mm} {\textstyle +\Theta \big(g_{up}g_{uq}g_{uu,r}+g_{uu,(p}g_{q)u}-g_{uu}g_{u(p}g_{q)u,r}-2g_{u(p}g_{q)u,u}-\frac{1}{2}g_{uu}g_{pq,u}\big)}
\,, \label{Wpq}
\end{eqnarray}
their contractions are defined as ${\,Q\equiv g^{pq}Q_{pq}\,}$, ${\,W \equiv g^{pq}W_{pq}\,}$ and ${\,X_q \equiv g^{pm}X_{pmq}\,}$, and the auxiliary quantities are defined in (\ref{auxiliaryB})--(\ref{auxiliaryE}).

As a generalization of the classical Petrov classification of four-dimensional spacetimes in Einstein's theory, classification scheme of the algebraic structure of the Weyl tensor is based on whether the scalars (\ref{defPsiCoef}) of various boost weights vanish or not in a \emph{suitable} null frame \cite{ColeyMilsonPravdaPravdova:2004}, see \cite{OrtaggioPravdaPravdova:2013} for a recent comprehensive review.

Specifically, it is possible to introduce the \emph{principal alignment types and subtypes} of the Weyl tensor in any dimension $D$ based on the existence of the (multiple) \emph{Weyl aligned null direction} (WAND)~$\boldk$, as summarized in table~\ref{PriClassSch}. Apart from a fully generic type G with all Weyl tensor components nonvanishing, there is type I with subtypes I(a) and I(b), type II with four possible subtypes II(a), II(b), II(c) and II(d), type III with two subtypes III(a) and III(b), type N and type O corresponding to the case when the Weyl tensor vanishes completely.

Subsequently, it is possible to introduce \emph{secondary alignment types} of the Weyl tensor defined by the property that there exists an \emph{additional} WAND~$\boldl$, namely $\mathrm{I}_i$, $\mathrm{II}_i$ and $\mathrm{III}_i$. These are the types I, II, and III, respectively, for which not only ${\Psi_{0^{ij}}=0}$ but also ${\Psi_{4^{ij}}=0}$. There is also the ``degenerate'' case of type D equivalent to $\mathrm{II}_{ii}$ for which only the zero-boost weight Weyl scalars $\Psi_{2^{...}}$ are nonvanishing.

\renewcommand{\arraystretch}{1.15}
\begin{table}[ht]
\begin{tabular}{c||lllll}
{\small type} & \multicolumn{5}{c}{{\small vanishing components}} \\ \hline\hline\hline
$\mathrm{G}$ & \multicolumn{5}{l}{{\small none (all $\Psi_{A^{...}}$ are nontrivial)}} \\ \hline\hline
$\mathrm{I}$ & $\Psi_{0^{ij}}$ & & & & \\ \hline
$\mathrm{I(a)}$ & $\Psi_{0^{ij}}$ & $\ \Psi_{1T^{i}}$ & & & \\
$\mathrm{I(b)}$ & $\Psi_{0^{ij}}$ & $\ \tilde{\Psi}_{1^{ijk}}$ & & & \\ \hline\hline
$\mathrm{II}$ & $\Psi_{0^{ij}}$ & $\ \Psi_{1T^{i}}$ $\tilde{\Psi}_{1^{ijk}}$ & & & \\ \hline
$\mathrm{II(a)}$ & $\Psi_{0^{ij}}$ & $\ \Psi_{1T^{i}}$ $\tilde{\Psi}_{1^{ijk}}$ & $\ \Psi_{2S}$ & & \\
$\mathrm{II(b)}$ & $\Psi_{0^{ij}}$ & $\ \Psi_{1T^{i}}$ $\tilde{\Psi}_{1^{ijk}}$ & $\ \tilde{\Psi}_{2T^{(ij)}}$ & & \\
$\mathrm{II(c)}$ & $\Psi_{0^{ij}}$ & $\ \Psi_{1T^{i}}$ $\tilde{\Psi}_{1^{ijk}}$ & $\ \tilde{\Psi}_{2^{ijkl}}$ & & \\
$\mathrm{II(d)}$ & $\Psi_{0^{ij}}$ & $\ \Psi_{1T^{i}}$ $\tilde{\Psi}_{1^{ijk}}$ & $\ \Psi_{2^{ij}}$ & & \\ \hline\hline
$\mathrm{III}$ & $\Psi_{0^{ij}}$ & $\ \Psi_{1T^{i}}$ $\tilde{\Psi}_{1^{ijk}}$ & $\ \Psi_{2S}$ $\tilde{\Psi}_{2T^{(ij)}}$ $\tilde{\Psi}_{2^{ijkl}}$ $\Psi_{2^{ij}}$ & & \\ \hline
$\mathrm{III(a)}$ & $\Psi_{0^{ij}}$ & $\ \Psi_{1T^{i}}$ $\tilde{\Psi}_{1^{ijk}}$ & $\ \Psi_{2S}$ $\tilde{\Psi}_{2T^{(ij)}}$ $\tilde{\Psi}_{2^{ijkl}}$ $\Psi_{2^{ij}}$ & $\ \Psi_{3T^{i}}$ & \\
$\mathrm{III(b)}$ & $\Psi_{0^{ij}}$ & $\ \Psi_{1T^{i}}$ $\tilde{\Psi}_{1^{ijk}}$ & $\ \Psi_{2S}$ $\tilde{\Psi}_{2T^{(ij)}}$ $\tilde{\Psi}_{2^{ijkl}}$ $\Psi_{2^{ij}}$ & $\ \tilde{\Psi}_{3^{ijk}}$ & \\ \hline\hline
$\mathrm{N}$ & $\Psi_{0^{ij}}$ & $\ \Psi_{1T^{i}}$ $\tilde{\Psi}_{1^{ijk}}$ & $\ \Psi_{2S}$ $\tilde{\Psi}_{2T^{(ij)}}$ $\tilde{\Psi}_{2^{ijkl}}$ $\Psi_{2^{ij}}$ & $\ \Psi_{3T^{i}}$ $\tilde{\Psi}_{3^{ijk}}$ & \\ \hline\hline
$\mathrm{O}$ & $\Psi_{0^{ij}}$ & $\ \Psi_{1T^{i}}$ $\tilde{\Psi}_{1^{ijk}}$ & $\ \Psi_{2S}$ $\tilde{\Psi}_{2T^{(ij)}}$ $\tilde{\Psi}_{2^{ijkl}}$ $\Psi_{2^{ij}}$ & $\ \Psi_{3T^{i}}$ $\tilde{\Psi}_{3^{ijk}}$ & $\ \Psi_{4^{ij}}$ \\
\end{tabular}
\caption{\label{PriClassSch} The principal alignment types and subtypes of the Weyl tensor defined by the existence of a (multiple) WAND $\boldk$.}
\end{table}
\begin{table}[ht]
\begin{tabular}{c||c|r}
{\small type} & {\small principal type} & {\small aditional vanishing components} \\ \hline\hline\hline
$\mathrm{I}_i$ & $\mathrm{I}$ & $\Psi_{4^{ij}}$ \\ \hline
$\mathrm{II}_i$ & $\mathrm{II}$ & $\Psi_{4^{ij}}$ \\ \hline
$\mathrm{III}_i$ & $\mathrm{III}$ & $\Psi_{4^{ij}}$ \\ \hline
$\mathrm{D}\equiv\mathrm{II}_{ii}$ & $\mathrm{II}$ & $\Psi_{3T^{i}}$ $\tilde{\Psi}_{3^{ijk}}$ $\Psi_{4^{ij}}$ \\
\end{tabular}
\caption{\label{SecClassSch} The secondary alignment types of the Weyl tensor defined by the existence of another WAND $\boldl$ (which is a double WAND in the case of type D).}
\end{table}

Of course, various combinations of these possibilities can occur. For example, there may be a spacetime with the algebraic structure II(ab) which means that it is \emph{both} of subtype II(a) and II(b). Clearly, II=I(ab), III=II(abcd), N=III(ab). In addition, there may be, for example, a $\mathrm{II(c)}_i$ spacetime which means that it is simultaneously of subtype II(c) and $\mathrm{II}_i$. Or there can be a D(bcd) spacetime defined by the property that there exists a double WAND $\boldk$ \emph{and} a double WAND $\boldl$ such that the only nonvanishing Weyl scalar is $\Psi_{2S}$.

In our present contribution, we are going to apply this algebraic classification scheme to the fully general family of non-twisting and shear-free geometries. This contains both the non-expanding Kundt class (${\Theta=0}$) and the expanding Robinson--Trautman class (${\Theta\neq 0}$). In particular, we will completely characterize all possible principal alignment types of the Weyl tensor with respect to the optically privileged null vector field ${\mathbf{k}=\mathbf{\partial}_r}$.

Using the fact that ${\Psi_{0^{ij}}=0}$, see (\ref{Psi0ij}), and table~\ref{PriClassSch}, it immediately follows that a \emph{generic Kundt or Robinson--Trautman geometry is of algebraic type} I (or more special), so that the optically privileged (non-twisting and shear-free) null vector field ${\mathbf{k}=\mathbf{\partial}_r}$ is the WAND.
Moreover, since the relation (\ref{Psi1ijk}) reads ${\tilde{\Psi}_{1^{ijk}}=0}$, \emph{any Kundt or Robinson--Trautman geometry is, in fact, of algebraic subtype} I(b), or more special.

We will now discuss all possibilities when ${\boldk=\mathbf{k}}$ is a \emph{multiple WAND}. In other words, the spacetime geometry is (at least) of algebraic type II with respect to $\mathbf{\partial}_r$.

\section{Multiple WAND ${\mathbf{k}}$ and algebraically special (sub)types}
\label{multipleWAND}

When the vector field ${\mathbf{k}=\mathbf{\partial}_r}$ is (at least) a \emph{double} WAND then the spacetime is \emph{algebraically special} (of type II, or more special).\footnote{In principle, there could exist ``peculiar'' algebraically special spacetimes for which ${\mathbf{k}}$ is \emph{not a double} WAND and there is \emph{another} double WAND vector field.} It follows from table~\ref{PriClassSch} that such a situation occurs if, and only if,
\begin{equation}
\Psi_{1T^{i}}=0 \,,
\end{equation}
where $\Psi_{1T^{i}}$ is given by (\ref{Psi1Tj}). Since the spatial vectors $\boldm_i$ are linearly independent, this condition is equivalent to ${\big(g_{up,r}-2\Theta g_{up}\big)_{,r}=2\Theta_{,p}}$ which can be rewritten as
\begin{equation}
g_{up,rr}=2g_{up}\Theta_{,r}+2\Theta g_{up,r}+2\Theta_{,p} \,, \label{podminka na I(a)=IIint}
\end{equation}
or integrated to
\begin{equation}
g_{up,r}=2\Theta g_{up}+f_p \,, \label{podminka na I(a)=II}
\end{equation}
where ${f_p(r,u,x)\equiv 2\int\Theta_{,p}\,\dd r+\varphi_p(u,x)}$, that is
\begin{equation}
f_{p,r}= 2\Theta_{,p} \,. \label{deff}
\end{equation}
Applying the condition ${\Psi_{1T^i}=0}$ which implies (\ref{podminka na I(a)=II}) and (\ref{podminka na I(a)=IIint}), the functions (\ref{P})--(\ref{Wpq}) determining the remaining Weyl scalars (\ref{Psi2s})--(\ref{Psi4ij}) simplify considerably to
\begin{eqnarray}
\fl P &=& {\textstyle \big(\frac{1}{2}g_{uu,r}-\Theta g_{uu}\big)_{,r}
+\frac{1}{(D-2)(D-3)}\,^{S}R } \nonumber\\
\fl && {\textstyle +\frac{1}{D-2}g^{mn}f_{m||n}
-\frac{1}{4}\frac{D-4}{D-2}g^{mn}f_mf_n-2\Theta_{,u}} \,, \label{PII}\\
\fl Q_{pq} &=& {\textstyle \,^{S}R_{pq}+\frac{1}{2}(D-4)\big(f_{(p||q)}+\frac{1}{2}f_pf_q
\big)}, \label{QpqII}\\
\fl F_{pq} &=& f_{[p,q]} \,, \label{FpqII}\\
\fl V_p &=& {\textstyle \frac{1}{2}\big(f_{p,u}-g_{uu,rp}-g_{up}g_{uu,rr}+g^{mn}f_mE_{np}\big)+\frac{1}{2}\frac{D-4}{D-3}g^{mn}g_{u[p}f_{m]}f_n}
\nonumber \\
\fl && {\textstyle -\frac{1}{D-3}\,g^{mn}\big[g_{m[p,u||n]}+g_{u[m,p]||n}+e_{m[p}f_{n]}
+\frac{1}{2}g_{up}f_{m||n}}\nonumber\\
\fl && \hspace{17.0mm} {\textstyle -\frac{1}{2}g_{um}(f_{(n||p)}-3f_{[n,p]})\big]} \nonumber \\
\fl && {\textstyle
+g_{up}g_{uu}\Theta_{,r}+2g_{up}\Theta_{,u}+g_{uu}\Theta_{,p}
+\Theta \big(g_{up}g_{uu,r}+g_{uu,p}\big)} , \label{VpII}\\
\fl X_{pmq}\!&=&{\textstyle g_{p[m,u||q]}+g_{u[q,m]||p}+e_{p[m}f_{q]}-g_{u[q}f_{m]||p}-g_{up}f_{[m,q]}-\frac{1}{2}g_{u[q}f_{m]}f_{p}} \,, \label{XpmqII}\\
\fl W_{pq}&=& {\textstyle -\frac{1}{2}g_{uu||p||q}-\frac{1}{2}g_{pq,uu}+g_{u(p,u||q)}-\frac{1}{2}g_{uu,r}e_{pq}-g_{uu,r(p}g_{q)u}-\frac{1}{2}g_{up}g_{uq}g_{uu,rr} }\nonumber\\
\fl && {\textstyle +\frac{1}{2}g_{uu,(p}f_{q)}+\frac{1}{2}g_{uu}f_{(p||q)}+g_{u(p}f_{q),u}
+\frac{1}{4}g^{mn}\big(g_{um}g_{un}f_{p}f_{q}+f_{m}f_{n}g_{up}g_{uq}\big) } \nonumber \\
\fl && {\textstyle -\frac{1}{2}g^{mn}g_{um}f_{n}g_{u(p}f_{q)}
+g^{mn}\big(E_{mp}E_{nq}+f_{m}E_{n(p}g_{q)u}-g_{um}E_{n(p}f_{q)}\big)} \nonumber \\
\fl && {\textstyle +g_{uu}g_{up}g_{uq}\Theta_{,r}+2g_{up}g_{uq}\Theta_{,u}+2g_{uu}g_{u(p}\Theta_{,q)}} \nonumber \\
\fl && {\textstyle +\Theta\big(2g_{uu,(p}g_{q)u}+g_{uu}g_{u(p||q)}+g_{up}g_{uq}g_{uu,r}-\frac{1}{2}g_{uu}g_{pq,u}\big)} . \label{WpqII}
\end{eqnarray}
Recall that
\begin{equation}
e_{pq} = g_{u{(p||q)}}- {\textstyle \frac{1}{2}}g_{pq,u} \,, \qquad
E_{pq} = g_{u{[p,q]}}+ {\textstyle \frac{1}{2}}g_{pq,u} \,, \label{epqEpqII}
\end{equation}
${f_{pq} = f_{(p||q)}+ \frac{1}{2}f_pf_q
+2g_{u(p}\Theta_{,q)}+2\Theta^2g_{up}g_{uq}+2\Theta\big(g_{u(p||q)}+g_{u(p}f_{q)}\big)}$,
and ${\,_{||}}$ denotes the covariant derivative with respect to the spatial metric $g_{pq}$, while the corresponding Ricci tensor and Ricci scalar are ${\,^{S}R_{pq}}$ and ${\,^{S}R}$, respectively, see~\ref{appendixA}.

Using table~\ref{PriClassSch} we can thus explicitly present the necessary and sufficient conditions determining all possible principal algebraic (sub)types of non-twisting shear-free geometries with a double, triple and quadruple WAND ${\mathbf{k}=\mathbf{\partial}_r}$. These are summarized in table~\ref{AlgSpecClassSch}.

\begin{table}[ht]
\begin{tabular}{c||r|c}
{\small type} & {\small necessary and sufficient condition} & {\small equation} \\ \hline\hline\hline
$\mathrm{II(a)}$ & ${P=0} \quad$ & (\ref{CondIIa})\\ \hline
$\mathrm{II(b)}$ & ${Q_{pq}-\frac{1}{D-2}\,g_{pq}\,Q=0} \quad$ & (\ref{CondIIb}) \\ \hline
$\mathrm{II(c)}$ & ${\,^{S}C_{mpnq}=0} \quad$ & (\ref{CondIIc}) \\ \hline
$\mathrm{II(d)}$ & ${F_{pq}=0} \quad$ & (\ref{CondIId}) \\ \hline\hline\hline
$\mathrm{III}$ & $\mathrm{II(abcd)} \qquad\qquad$ & \\ \hline\hline
$\mathrm{III(a)}$ & ${V_p=0} \quad$ & (\ref{CondIIIa}) \\ \hline
$\mathrm{III(b)}$ & ${X_{pmq}-\frac{2}{D-3}\,g_{p[m}X_{q]}=0} \quad$ & (\ref{CondIIIb}) \\ \hline\hline\hline
$\mathrm{N}$ & $\mathrm{III(ab)} \qquad\qquad$ & \\ \hline\hline\hline
$\mathrm{O}$ & ${W_{pq}-\frac{1}{D-2}\,g_{pq}\,W=0} \quad$ & (\ref{CondO})\\
\end{tabular}
\caption{\label{AlgSpecClassSch} Principal alignment types and subtypes of the algebraically special Weyl tensor with respect to the multiple WAND ${\mathbf{k}=\mathbf{\partial}_r}$.}
\end{table}

\subsection{Type II subtypes with a double WAND $\mathbf{k}$}
\label{subtypesofII}

The Robinson--Trautman or Kundt spacetimes (\ref{general nontwist}) with (\ref{shearfree condition}) satisfying the condition (\ref{podminka na I(a)=II}), i.e., ${\Psi_{1T^j}=0}$ implying that they are (at least) of type II with respect to the null direction ${\mathbf{k}=\mathbf{\partial}_r}$, admit the following particular algebraic subtypes of the Weyl tensor:

\smallskip
\begin{itemize}
\item \emph{subtype} II(a) $\Leftrightarrow$ ${\Psi_{2S}=0}$ $\Leftrightarrow$ ${P=0}$ $\Leftrightarrow$ the metric function $g_{uu}$ satisfies the relation:
\begin{eqnarray}
&& \fl {\textstyle \big(\frac{1}{2}g_{uu,r}-\Theta g_{uu}\big)_{,r}=-\frac{1}{(D-2)(D-3)}\,^{S}R
-\frac{1}{D-2}g^{mn}f_{m||n} + \frac{1}{4}\frac{D-4}{D-2}g^{mn}f_mf_n
+2\Theta_{,u}} \,.\nonumber\\
&& \label{CondIIa}
\end{eqnarray}
This determines the specific dependence of $g_{uu}(r,u,x)$ on the coordinate $r$ which is the affine parameter along the null congruence generated by $\mathbf{k}$.

\item \emph{subtype} II(b) $\Leftrightarrow$ ${\tilde{\Psi}_{2T^{(ij)}}=0}$ $\Leftrightarrow$ ${Q_{pq}=\frac{1}{D-2}\,g_{pq}\,Q}$:
\begin{eqnarray}
&& \fl \,^{S}R_{pq}-\frac{g_{pq}}{D-2}\,^{S}R
=-{\textstyle \frac{1}{2}}(D-4)
\Big[\!\big(f_{(p||q)}+{\textstyle \frac{1}{2}}f_pf_q\big)
-\frac{g_{pq}}{D-2}g^{mn}\big(f_{m||n}+{\textstyle \frac{1}{2}}f_mf_n
\big)\!\Big]. \nonumber\\
&& \label{CondIIb}
\end{eqnarray}
This is identically satisfied when ${D=4}$ since for any 2-dimensional Riemannian space there is ${\,^{S}\!R_{pq}=\frac{1}{2}g_{pq}\,^{S}\!R}$.

\item \emph{subtype} II(c) $\Leftrightarrow$ ${\tilde{\Psi}_{2^{ijkl}}=0}$:
\begin{equation}
\,^{S}C_{mpnq}=0 \,. \label{CondIIc}
\end{equation}
This is always satisfied when ${D=4}$ and ${D=5}$ since the Weyl tensor vanishes identically in dimensions 2 and 3.

\item \emph{subtype} II(d) $\Leftrightarrow$ ${\Psi_{2^{ij}}=0}$ $\Leftrightarrow$ ${F_{pq}=0}$:
\begin{equation}
f_{[p,q]}=0 \,. \label{CondIId}
\end{equation}
Introducing a 1-form ${\df\equiv f_p\,\dd x^p}$ in the transverse ${(D-2)}$-dim Riemannian space, this condition is equivalent to the condition that $\df$ is closed (${\dd \df=0}$). By the Poincar\'{e} lemma, on any contractible domain there exists a potential function ${\cal F}$ such that ${\df=\dd{\cal F}}$, that is ${f_p={\cal F}_{,p}}$.
In a general case, such ${\cal F}$ exists only \emph{locally}.
\end{itemize}
\smallskip

\noindent
These four distinct subtypes of type II can be arbitrarily combined. Clearly, in the ${D=4}$ case the algebraically special non-twisting shear-free geometries are always of subtype II(bc).

\subsection{Type III subtypes with a triple WAND $\mathbf{k}$}
\label{subtypesofIII}

The Robinson--Trautman or Kundt spacetime is of algebraic type~III with respect to the triple WAND ${\mathbf{k}=\partial_r}$ if \emph{all four} independent conditions (\ref{CondIIa})--(\ref{CondIId}) are satisfied \emph{simultaneously}. In such a case the zero-boost-weight Weyl tensor components $\Psi_{2^{...}}$ vanish and we obtain geometries of type II(abcd)$\equiv$III, with the remaining Weyl scalars (\ref{Psi3Tj})--(\ref{Psi4ij}) determined by the structural functions (\ref{VpII})--(\ref{WpqII}) now simplified to
\begin{eqnarray}
\fl V_p &=& {\textstyle \frac{1}{2}\big(f_{p,u}-g_{uu,rp}+g^{mn}f_mE_{np}-\frac{1}{2}g^{mn}g_{um}f_{n}f_p\big)+X_p} \nonumber \\
\fl && {\textstyle +\frac{1}{D-2}\,g_{up}\big[\frac{1}{D-3}\,^{S}R+g^{mn}\big(f_{m||n}+\frac{1}{2}f_mf_n\big)\big]
+g_{uu}\Theta_{,p}+\Theta g_{uu,p}} \,, \label{VpIII}\\
\fl X_{pmq}\!&=&{\textstyle g_{p[m,u||q]}+g_{u[q,m]||p}+e_{p[m}f_{q]}-g_{u[q}f_{m]||p}-\frac{1}{2}g_{u[q}f_{m]}f_{p}} \,, \label{XpmqIII}\\
\fl W_{pq}&=& {\textstyle -\frac{1}{2}g_{uu||p||q}-\frac{1}{2}g_{pq,uu}+g_{u(p,u||q)}-\big(\frac{1}{2}g_{uu,r}-\Theta g_{uu}\big)e_{pq}-g_{uu,r(p}g_{q)u}} \nonumber \\
\fl && {\textstyle +\frac{1}{2}g_{uu,(p}f_{q)}+\frac{1}{2}g_{uu}f_{(p||q)}+g_{u(p}f_{q),u}
+\frac{1}{4}g^{mn}g_{um}g_{un}f_{p}f_{q} } \nonumber \\
\fl && {\textstyle
-\frac{1}{2}g^{mn}g_{um}f_{n}g_{u(p}f_{q)}
+g^{mn}\big(E_{mp}E_{nq}+f_{m}E_{n(p}g_{q)u}-g_{um}E_{n(p}f_{q)}\big)} \nonumber \\
\fl && {\textstyle +\frac{1}{D-2}\,g_{up}g_{uq}\big[\frac{1}{D-3}\,^{S}R+g^{mn}\big(f_{m||n}+\frac{1}{2}f_mf_n\big)
\big]} \nonumber\\
\fl && {\textstyle +2g_{uu}g_{u(p}\Theta_{,q)}+2\Theta g_{uu,(p}g_{q)u}} \,. \label{WpqIII}
\end{eqnarray}
Consequently, in view of table~\ref{PriClassSch}, the explicit conditions for the subtypes III(a) and III(b) with the triple WAND~$\mathbf{k}$ are:

\smallskip
\begin{itemize}
\item \emph{subtype} III(a) $\Leftrightarrow$ ${\Psi_{3T^i}=0}$ $\Leftrightarrow$ ${V_p=0}$ $\Leftrightarrow$ the function $g_{uu,rp}$ is explicitly given as:
\begin{eqnarray}
g_{uu,rp} &=& {\textstyle f_{p,u}+g^{mn}f_mE_{np}-\frac{1}{2}g^{mn}g_{um}f_{n}f_p
+\frac{2}{D-3}X_p} \nonumber \\
&& {\textstyle +\frac{2}{D-2}\,g_{up}\big[\frac{1}{D-3}\,^{S}R+g^{mn}\big(f_{m||n}+\frac{1}{2}f_mf_n\big)\big]} \nonumber \\
&& {\textstyle +2g_{uu}\Theta_{,p}+2\Theta g_{uu,p}} \,. \label{CondIIIa}
\end{eqnarray}
This is a specific restriction on the spatial derivatives of the function $g_{uu,r}$.

\item \emph{subtype} III(b) $\Leftrightarrow$ ${\tilde{\Psi}_{3^{ijk}}=0}$:
\begin{equation}
X_{pmq}={\textstyle\frac{2}{D-3}}\,g_{p[m}X_{q]}\, \quad\hbox{where}\quad X_q = g^{pm}X_{pmq}\,, \label{CondIIIb}
\end{equation}
and $X_{pmq}$ is given by expression (\ref{XpmqIII}). Using the fact that any 2-dimensional metric $g_{pq}$ is conformally flat, ${g_{pq}=\Omega\, \delta_{pq}}$, it can be easily checked that the condition (\ref{CondIIIb}) is identically satisfied in ${D=4}$.

\end{itemize}

\subsection{Type N with a quadruple WAND $\mathbf{k}$}
\label{typeN}

When \emph{both} conditions (\ref{CondIIIa}) and (\ref{CondIIIb}) are satisfied, the only remaining Weyl scalar is
${\Psi_{4^{ij}} = m_i^pm_j^q\,\big(W_{pq}-\frac{1}{D-2}\,g_{pq} W\big)}$, see (\ref{Psi4ij}). In such a case we obtain the Robinson--Trautman or Kundt spacetimes of algebraic type~N with the quadruple WAND ${\mathbf{k}=\partial_r}$. Using ${V_p=0}$, that is by substituting (\ref{CondIIIa}) into (\ref{WpqIII}), the function ${W_{pq}}$ for such type N geometries reduces to a simple expression
\begin{eqnarray}
W_{pq}&=& {\textstyle -\frac{1}{2}g_{uu||p||q}-\frac{1}{2}g_{pq,uu}+g_{u(p,u||q)}-\big(\frac{1}{2}g_{uu,r}-\Theta\,g_{uu}\big)e_{pq}}\nonumber\\
&& {\textstyle +\frac{1}{2}g_{uu,(p}f_{q)}+\frac{1}{2}g_{uu}f_{(p||q)}+\frac{1}{4}g^{mn}g_{um}g_{un}f_{p}f_{q} } \nonumber \\
&& {\textstyle -\frac{1}{D-2}\,g_{up}g_{uq}\big[\frac{1}{D-3}\,^{S}R+g^{mn}\big(f_{m||n}+\frac{1}{2}f_mf_n\big)\big]} \nonumber \\
&& {\textstyle
-\frac{2}{D-3}\,X_{(p}g_{q)u}+g^{mn}\big(E_{mp}E_{nq}-g_{um}E_{n(p}f_{q)}\big)}\,, \label{WpqN}
\end{eqnarray}
determined by the metric functions (\ref{general nontwist}) and their first and second derivatives. The set of functions $W_{pq}$ directly encodes the amplitudes $\Psi_{4^{ij}}$ of the corresponding gravitational waves, forming a symmetric traceless matrix of dimension ${(D-2)\times(D-2)}$.

\subsection{Type O geometries}
\label{typeO}

The Weyl tensor vanishes completely if, and only if, \emph{all} the above conditions are satisfied and, \emph{in addition\,}, ${\Psi_{4^{ij}}=0}$. This clearly occurs when
\begin{equation}
W_{pq}={\textstyle\frac{1}{D-2}}\,g_{pq} W\,, \label{CondO}
\end{equation}
with ${W=g^{pq}W_{pq}}$, which is a restriction on the functions $W_{pq}$ given by (\ref{WpqN}).

\section{Secondary alignment types of the Weyl tensor}
\label{otherWANDl}

In the non-twisting and shear-free geometries there may also exist an \emph{additional} WAND~$\boldl$ distinct from the (multiple) WAND ${\boldk=\mathbf{k}=\mathbf{\partial}_r}$. In such cases the spacetimes are of type $\mathrm{I}_i$, $\mathrm{II}_i$, $\mathrm{III}_i$ or D, see table~\ref{sec_geom}.

In general, when ${\Psi_{4^{ij}} = m_i^pm_j^q\,\big(W_{pq}-\frac{1}{D-2}\,g_{pq} W\big)=0}$ with $W_{pq}$ given by (\ref{Wpq}), the geometry is of type $\mathrm{I}_i$ (the subtype $\mathrm{I}(b)_i$, in fact). In such a case ${\boldk=\mathbf{k}=\mathbf{\partial}_r}$ and ${\boldl=\frac{1}{2}g_{uu}\mathbf{\partial}_r+\mathbf{\partial}_u}$ are two distinct WANDs, see~(\ref{nat null frame}).

When the geometry is of type II with the double WAND $\boldk$ \emph{and} ${\Psi_{4^{ij}} =0}$ where $W_{pq}$ is given by (\ref{WpqII}), the geometry is of type $\mathrm{II}_i$ with another WAND $\boldl$. If the additional conditions (\ref{CondIIa}), (\ref{CondIIb}), (\ref{CondIIc}) and (\ref{CondIId}) are satisfied, we obtain the subtypes $\mathrm{II(a)}_i$, $\mathrm{II(b)}_i$, $\mathrm{II(c)}_i$ and $\mathrm{II(d)}_i$, respectively (or their combinations).

The type $\mathrm{III}_i$ geometry is equivalent to $\mathrm{II(abcd)}_i$, in which case there is the triple WAND $\boldk$ and an additional WAND $\boldl$. The subtypes $\mathrm{III(a)}_i$ and $\mathrm{III(b)}_i$ occur if conditions (\ref{CondIIIa}) and (\ref{CondIIIb}) are also satisfied. For such spacetimes, the only non-vanishing Weyl scalars are $\tilde{\Psi}_{3^{ijk}}$ and $\Psi_{3T^i}$, respectively.

Finally, there is also the ``degenerate'' case $\mathrm{D}\equiv\mathrm{II}_{ii}$ which admits the \emph{double} WAND ${\boldk=\mathbf{\partial}_r}$ and the \emph{double} WAND ${\boldl=\frac{1}{2}g_{uu}\mathbf{\partial}_r+\mathbf{\partial}_u}$. Its only nonvanishing Weyl scalars are $\Psi_{2^{...}}$, i.e., those of zero-boost-weight. In view of (\ref{Psi0ij})--(\ref{Psi4ij}) and (\ref{PII})--(\ref{WpqII}), this occurs if (and only if) ${V_p=0}$, ${X_{pmq}=\frac{2}{D-3}\,g_{p[m}X_{q]}}$, ${W_{pq}=\frac{1}{D-2}\,g_{pq}\,W}$, in which the functions are determined by (\ref{VpII}), (\ref{XpmqII}), (\ref{WpqII}). Of course, with the additional constraints (\ref{CondIIa}), (\ref{CondIIb}), (\ref{CondIIc}) and (\ref{CondIId}) we obtain the subtypes D(a), D(b), D(c) and D(d), respectively, and their various combinations. The simplest type D geometry thus seems to be of the subtype D(bcd) for which the only nonvanishing Weyl scalar is ${\Psi_{2S}= \frac{D-3}{D-1}\, P}$ with $P$ given by expression (\ref{PII}). This involves, for example, generalizations of Schwarzschild black hole spacetimes, see subsection~\ref{EinsteinRTexample}.

\section{Kundt geometries}
\label{Kundt_geometries}

We will now discuss the two distinct important subclasses of non-twisting shear-free geometries, namely the non-expanding Kundt and the expanding Robinson--Trautman metrics (in the next section).

The Kundt family is defined by having a \emph{vanishing expansion}, ${\Theta=0}$. It implies that the spatial metric ${g_{pq}}$ in (\ref{general nontwist}) is $r$-independent,
\begin{equation}
g_{pq}\equiv h_{pq}(u,x) \,, \label{Kundtgpq}
\end{equation}
see end of section~\ref{sec_geom}. This significantly simplifies the Riemann tensor (\ref{Riem rprq})--(\ref{Riem upuq}), the Ricci tensor (\ref{Ricci rr})--(\ref{Ricci uu}) and the Weyl tensor (\ref{Weyl rprq})--(\ref{Weyl upuq}) listed explicitly in~\ref{appendixA}. In fact, it is a complete generalization of the analogous results presented previously in~\cite{PodolskySvarc:2013a} since \emph{no field equations} and \emph{no constraints on algebraically special types} have been employed. The curvature tensor components given \hbox{in~\ref{appendixA}} thus characterize the most general Kundt geometry, which is of algebraic type~I(b), whereas the results in Appendix~B of~\cite{PodolskySvarc:2013a} are only valid for Kundt spacetimes of type~II (or more special).

For such a generic Kundt geometry the Weyl tensor frame components are given by (\ref{Psi0ij})--(\ref{Psi4ij}) with (\ref{P})--(\ref{Wpq}), simplified by setting ${\Theta=0}$. These represent an explicit form of the expressions (7)--(16) written in~\cite{PodolskySvarc:2013a}.

In the case of a vanishing expansion we may fully express the conditions discussed in section~\ref{multipleWAND} for algebraically special Kundt geometries with respect to the WAND ${\mathbf{k}=\mathbf{\partial}_r}$. Specifically, we integrate equations (\ref{podminka na I(a)=II}) and (\ref{CondIIa}), obtaining an explicit $r$-dependence of the metric functions $g_{up}$ and $g_{uu}$, respectively. After substituting them into the remaining conditions and separating in $r$, we obtain:

\smallskip
\begin{itemize}
\item
The type I Kundt geometry is of subtype~I(a)$=$I(ab)$\equiv$II ${\Leftrightarrow{\Psi_{1T^i}=0\,} \Leftrightarrow}$
\begin{equation}
g_{up}=e_p(u,x)+ f_p(u,x) \,r\,, \label{KundtII}
\end{equation}
where $e_p$ and $f_p$ are arbitrary functions of the coordinates $u$ and $x$.

\item
The type II$\equiv$I(ab) Kundt geometry is of subtype~II(a) ${\Leftrightarrow\Psi_{2S} =0 \Leftrightarrow}$
\begin{equation}
g_{uu}=a(u,x)\,r^2+ b(u,x)\,r+c(u,x)\,, \label{KundtIIa}
\end{equation}
where ${a=\frac{1}{4}f^p f_p-\frac{1}{D-2}\big(\frac{1}{D-3}\,^{S}\!R+g^{pq}f_{pq}\big)}$ with ${f^p\equiv g^{pq}f_q}$, ${f_{pq}\equiv f_{(p||q)}+\frac{1}{2}f_pf_q}$, $b$~and~$c$ are arbitrary functions of the coordinates $u$ and $x$.

\item
The type II Kundt geometry is of subtype~II(b) ${\Leftrightarrow\tilde\Psi_{2T^{(ij)}} =0 \Leftrightarrow}$
\begin{equation}
\,^{S}\!R_{pq}-\frac{g_{pq}}{D-2}\,^{S}\!R=
-{\textstyle\frac{1}{2}}(D-4)\Big(f_{pq}-\frac{g_{pq}}{D-2}\,g^{mn}f_{mn}\Big)\,. \label{KundtIIb}
\end{equation}

\item
The type II Kundt geometry is of subtype~II(c) ${\Leftrightarrow\tilde\Psi_{2^{ijkl}}=0 \Leftrightarrow}$
\begin{equation}
\,^{S}\!C_{mpnq}=0 \,. \label{KundtIIc}
\end{equation}

\item
The type II Kundt geometry is of subtype~II(d) ${\Leftrightarrow\Psi_{2^{ij}} =0 \Leftrightarrow}$
\begin{equation}
F_{pq}\equiv f_{[p,q]}=0\,. \label{KundtIId}
\end{equation}

\item
The type III$\equiv$II(abcd) Kundt geometry is of subtype~III(a) ${\Leftrightarrow{\Psi_{3T^i}=0} \Leftrightarrow}$
\begin{eqnarray}
a_{,q}+f_{q}\,a = 0 \,, \qquad b_{,q}-f_{q,u} = T_q \,, \label{KundtIIIa}
\end{eqnarray}
where
\begin{eqnarray}
T_q \equiv {\textstyle -2\,e_{q}\big(a-\frac{1}{4}f^p f_p\big)-\frac{1}{2}e^pf_pf_q+f^p E_{pq}+\frac{2}{D-3}\,X_q} \,, \label{defTq}
\end{eqnarray}
with ${e^p\equiv g^{pq}e_q}$, ${e_{pq}^{_\mathrm{K}}\equiv e_{(p||q)}-\frac{1}{2}g_{pq,u}\,}$, ${E_{pq}= e_{[p,q]}+\frac{1}{2}g_{pq,u}\,}$, ${X_q\equiv g^{pm}X_{pmq}}$ and
\begin{equation}
X_{pmq}={\textstyle g_{p[m,u||q]}+e_{[q,m]||p}+e^{_\mathrm{K}}_{p[m}f_{q]}-e_{[q}f_{m]p}} \,.
\end{equation}
\item
The type~III Kundt geometry is of subtype~III(b) ${\Leftrightarrow{\tilde\Psi_{3^{ijk}}=0} \Leftrightarrow}$
\begin{equation}
{\textstyle X_{pmq}=\frac{1}{D-3}\big( g_{pm}\,X_q-g_{pq}\,X_m\big)} \,. \label{KundtIIIb}
\end{equation}

\item The type~N$\equiv$III(ab) Kundt geometry is completely described by the symmetric traceless matrix ${\Psi_{4^{ij}}}$, which is determined by
\begin{eqnarray}
\fl W_{pq} &=& {\textstyle
-\frac{1}{2}c_{||p||q}+\frac{1}{2}c_{,(p}f_{q)}+\frac{1}{2}cf_{(p||q)}
-\frac{1}{2}b\,e_{pq}^{_\mathrm{K}} +\big(a-\frac{1}{4}f^nf_n\big)e_{p}e_{q}+e_{(p,u||q)}} \nonumber \\
\fl && {\textstyle -\frac{1}{2}\,g_{pq,uu}+\frac{1}{4}e^{n}e_{n}\,f_{p}f_{q}-e^{n}E_{n(p}f_{q)} +g^{mn}E_{mp}E_{nq}-\frac{2}{D-3}\,X_{(p}\,e_{q)}}\nonumber\\
\fl && {\textstyle
-\frac{1}{2}}\big(2a\,e_{pq}^{_\mathrm{K}}
+T_{(p||q)}+T_{(p}f_{q)}-f_{(p,u||q)}-f_{(p} f_{q),u}\big)\,r \,. \label{KundtNWpq}
\end{eqnarray}
\item The type~N becomes type~O ${\Leftrightarrow{\Psi_{4^{ij}}=0}\Leftrightarrow W_{pq}=\frac{1}{D-2}\, g_{pq}\,W}$ with ${W\equiv g^{pq}W_{pq}}$.
\end{itemize}
\newpage

\noindent
These are the same conditions as those presented in~\cite{PodolskySvarc:2013a}. For Kundt geometries of type II (whose metric functions $g_{up}$ satisfy the condition (\ref{KundtII}))
the shorthands (\ref{FpqII}), (\ref{epqEpqII}), (\ref{XpmqII}) used in this paper become
\begin{eqnarray}
F_{pq} &=& f_{[p||q]} \,, \\
f_{pq} &=& f_{(p||q)}+{\textstyle \frac{1}{2}}f_{p}f_{q} \,, \\
E_{pq} &=& e_{[p||q]}+{\textstyle \frac{1}{2}}g_{pq,u}+r\,f_{[p||q]}\,, \\
e_{pq} &=& e_{(p||q)}-{\textstyle \frac{1}{2}}g_{pq,u}+r\,f_{(p||q)}\,, \\
X_{pmq} &=& e_{[q||m]||p}+F_{qm}e_p + F_{p[m}e_{q]}+e_{p[m}f_{q]}-f_{p[m}e_{q]}+g_{p[m,u||q]} \nonumber \\
&& +r\,\big(f_{[q||m]||p}+ F_{qm}f_p + F_{p[m}f_{q]}\big)\,.
\end{eqnarray}
The identification is
\begin{eqnarray}
F_{pq} &\equiv& F_{pq}^{_\mathrm{K}} \,, \\
f_{pq} &\equiv& f_{pq}^{_\mathrm{K}} \,, \\
E_{pq} &\equiv& E_{pq}^{_\mathrm{K}}+r\,F_{pq}^{_\mathrm{K}} \,, \\
e_{pq} &\equiv& e_{pq}^{_\mathrm{K}}+r\,f_{(p||q)} \,, \\
X_{pmq}&\equiv& X^{_\mathrm{K}}_{pmq}+r\,Y^{_\mathrm{K}}_{pmq} \,,
\end{eqnarray}
where the superscript $\,^{_\mathrm{K}}$ denotes the quantities defined and employed in \cite{PodolskySvarc:2013a}.

\section{Robinson--Trautman geometries}
\label{RT_geometries}

The algebraic structure of generic Robinson--Trautman geometries with an arbitrary ${\Theta\neq 0}$ has been described in sections~\ref{genralPsi} and \ref{multipleWAND}. Let us now investigate in detail a large particular subclass such that the non-twisting shear-free congruence generated by the null vector field ${\mathbf{k}=\mathbf{\partial}_r}$ has an expansion of the form
\begin{equation}
\Theta=\frac{1}{r} \,. \label{Theta_je_1/r}
\end{equation}
This is an important subcase since ${\Theta_{,r}+\Theta^2=0}$ and thus, in view of
(\ref{Riem rprq}) and (\ref{Ricci rr}), there is ${R_{rprq} = 0 = R_{rr}}$. Consequently, ${R_{abcd}\; k^a\, m_i^b\, k^c\, m_j^d = 0 = R_{ab}\; k^a k^b }$ which means that \emph{such Robinson--Trautman geometries are of Riemann type I and also of Ricci type I}.

For the case (\ref{Theta_je_1/r}) we can explicitly integrate all the conditions with respect to $r$ and determine the algebraic types and subtypes. First, the spatial metric $g_{pq}$ becomes
\begin{equation}
g_{pq}=r^2\,h_{pq}(u,x) \,, \label{RTgpq}
\end{equation}
see (\ref{IntShearFreeCond}) for ${R=r}$, which is obtained by solving ${R_{,r}=\Theta R\,}$. Such geometries are, in general, of the Weyl (sub)type I(b).

The metrics are of the Weyl type II (or more special) with the double WAND~${\mathbf{k}}$ if, and only if, the condition ${\Psi_{1T^i}=0}$ is satisfied. For (\ref{Theta_je_1/r}) we have ${f_{p,r}=0}$ due to (\ref{deff}), i.e., the functions $f_p$ are \emph{independent of} $r$. By integrating (\ref{podminka na I(a)=II}) we then obtain
\begin{equation}
g_{up}= e_p(u,x)\,r^2-f_p(u,x)\,r \,, \label{RTgup}
\end{equation}
where $e_p$ and $f_p$ are arbitrary functions of the coordinates $u$ and $x$.

With the conditions (\ref{Theta_je_1/r})--(\ref{RTgup}), the functions (\ref{PII})--(\ref{WpqII}) determining the Weyl scalars reduce to
\begin{eqnarray}
\fl P &=& {\textstyle \big(\frac{1}{2}g_{uu,r}- g_{uu}\,r^{-1} \big)_{,r}
-\alpha\,r^{-2}} \,, \label{PIIRT}\\
\fl Q_{pq} &=& {\textstyle \mathcal{R}_{pq}+\frac{1}{2}(D-4)\big(f_{(p||q)}+\frac{1}{2}f_pf_q
\big)}, \label{QpqIIRT}\\
\fl F_{pq} &=& f_{[p,q]} \,, \label{FpqIIRT}\\
\fl V_p &=& {\textstyle -\frac{1}{2}g_{uu,rp}+g_{uu,p}\,r^{-1}-\big(e_p\,r^2-f_p\,r\big)\big(\frac{1}{2}g_{uu,rr}-g_{uu,r}\,r^{-1}+g_{uu}\,r^{-2}\big)} \nonumber \\
\fl && {\textstyle +\frac{1}{2}f_{p,u}+\frac{1}{2}f^nE^{_\mathrm{RT}}_{np}+\frac{1}{2}\frac{D-4}{D-3}f^{n}e_{[p}f_{n]}-\frac{1}{2}\frac{D-6}{D-3}f^{n}F_{np}\,r^{-1}} \nonumber \\
\fl && {\textstyle -\frac{1}{D-3}\,h^{mn}\big[h_{m[p,u||n]}+e_{[m,p]||n}+e^{_\mathrm{RT}}_{m[p}f_{n]}} \nonumber \\
\fl && \hspace{18.0mm}{\textstyle +\frac{1}{2}\big(e_pf_{m||n}-e_mf_{(n||p)}\big)+\frac{3}{2}e_mF_{np}-F_{mp||n}\,r^{-1}\big]} \,, \label{VpIIRT}\\
\fl X_{pmq}\!&=& X^{_\mathrm{RT}}_{pmq}\,r^2+ Y^{_\mathrm{RT}}_{pmq}\,r \,, \label{XpmqIIRT}\\
\fl W_{pq}&=& {\textstyle -\frac{1}{2}g_{uu||p||q}+e_{pq}^{_\mathrm{RT}}\big(g_{uu}\,r-\frac{1}{2}g_{uu,r}\,r^2\big)-\frac{1}{2}f_{(p||q)}\big(g_{uu}-g_{uu,r}\,r\big)}\nonumber\\
\fl && {\textstyle -g_{uu,r(p}e_{q)}\,r^2+2g_{uu,(p}e_{q)}\,r-\frac{3}{2}g_{uu,(p}f_{q)}+g_{uu,r(p}f_{q)}\,r} \nonumber \\
\fl && {\textstyle -\big(e_pe_qr^2-2e_{(p}f_{q)}\,r+f_pf_q\big)\big(\frac{1}{2}g_{uu,rr}\,r^2-g_{uu,r}\,r+g_{uu}\big)} \nonumber \\
\fl && {\textstyle +\big[e_{(p,u||q)}+e_{(p}f_{q),u}-\frac{1}{2}h_{pq,uu}+\frac{1}{4}\big(e^ne_nf_{p}f_{q}+f^{n}f_{n}e_{p}e_{q}\big)-\frac{1}{2}e^nf_ne_{(p}f_{q)}} \nonumber \\
\fl && \hspace{5.6mm} {\textstyle +h^{mn}E^{_\mathrm{RT}}_{mp}E^{_\mathrm{RT}}_{nq}+f^nE^{_\mathrm{RT}}_{n(p}e_{q)}-e^nE^{_\mathrm{RT}}_{n(p}f_{q)} \big]}\,r^2 \nonumber \\
\fl && {\textstyle -\big[f_{(p,u||q)}+f_{(p}f_{q),u}-2h^{mn}E^{_\mathrm{RT}}_{m(p}F_{q)n}+f^{n}F_{n(p}e_{q)}-e^{n}F_{n(p}f_{q)} \big]}\,r \nonumber \\
\fl && {\textstyle +h^{mn}F_{mp}\,F_{nq}} \,. \label{WpqIIRT}
\end{eqnarray}
where ${\mathcal{R}=\,^{S}\!R\,r^2}$ and ${\mathcal{R}_{pq}=\!\,^{S}\!R_{pq}}$ are the Ricci scalar and the Ricci tensor with respect to the Riemannian metric ${h_{pq}(u,x)}$, respectively,
\begin{eqnarray}
\fl \alpha(u,x) \!&\equiv&{\textstyle \frac{1}{4}f^pf_p-\frac{1}{D-2}\big(\frac{1}{D-3}\,\mathcal{R}+{f^p}_{||p}+\frac{1}{2}f^pf_p\big)}\,, \label{alphaIIRT}\\
\fl X^{_\mathrm{RT}}_{pmq}\!&\equiv&{\textstyle h_{p[m,u||q]}+e_{[q,m]||p}+e^{_\mathrm{RT}}_{p[m}f_{q]}-e_{[q}f_{m]||p}-e_{p}F_{mq}-\frac{1}{2}e_{[q}f_{m]}f_{p}} \,, \label{XpmqIIRTRT}\\
\fl Y^{_\mathrm{RT}}_{pmq}\!&\equiv&{\textstyle -F_{qm||p}+f_{[q}F_{m]p}+f_{p}F_{mq}}\,, \label{YpmqIIRTRT}
\end{eqnarray}
and
\begin{eqnarray}
e^p\ \erovno h^{pq}e_q \,, \hspace{23mm}
f^p\ \equiv h^{pq}f_q \,, \\
e^{_\mathrm{RT}}_{pq} \erovno e_{(p||q)}-{\textstyle\frac{1}{2}}h_{pq,u}\,, \hspace{7.3mm}
E^{_\mathrm{RT}}_{pq} \equiv e_{[p||q]}+{\textstyle\frac{1}{2}}h_{pq,u}\,, \\
X^{_\mathrm{RT}}_q \erovno h^{pm}X^{_\mathrm{RT}}_{pmq} \,,\hspace{15.4mm}
Y^{_\mathrm{RT}}_q \equiv h^{pm}Y^{_\mathrm{RT}}_{pmq} \,. \label{XYRTq}
\end{eqnarray}
Using the results of subsections \ref{subtypesofII}--\ref{typeO} we can thus explicitly express the conditions for the principal alignment (sub)types of the algebraically special Weyl tensor:

\smallskip
\begin{itemize}
\item
The type II Robinson--Trautman geometry is of subtype~II(a) ${\Leftrightarrow\Psi_{2S} =0 \Leftrightarrow}$
\begin{equation}
g_{uu}= \alpha(u,x)+ \beta(u,x)\,r+\gamma(u,x)\,r^2\,, \label{RTIIa}
\end{equation}
where $\alpha$ is given by (\ref{alphaIIRT}) while $\beta$, $\gamma$ are arbitrary func\-tions of $u$ and $x$.

\item
The type II Robinson--Trautman geometry is of subtype~II(b) ${\Leftrightarrow\tilde\Psi_{2T^{(ij)}} =0 \Leftrightarrow}$
\begin{equation}
\fl \mathcal{R}_{pq}-\frac{h_{pq}}{D-2}\,\mathcal{R}=
-{\textstyle\frac{1}{2}}(D-4)\Big[\big(f_{(p||q)}+{\textstyle\frac{1}{2}}f_pf_q\big)-\frac{h_{pq}}{D-2}\big({f^n}_{||n}+{\textstyle\frac{1}{2}}f^nf_n\big)\Big], \label{RTIIb}
\end{equation}
where $\mathcal{R}_{pq}$ is the Ricci tensor with respect to the metric ${h_{pq}}$.

\item
The type II Robinson--Trautman geometry is of subtype~II(c) ${\Leftrightarrow\tilde\Psi_{2^{ijkl}}=0 \Leftrightarrow}$
\begin{equation}
\mathcal{C}_{mpnq}=0 \,, \label{RTIIc}
\end{equation}
where ${\mathcal{C}_{mpnq}=\,\!^{S}\!C_{mpnq}\,r^{-2}}$ is the Weyl tensor corresponding to the metric~${h_{pq}}$.

\item
The type II Robinson--Trautman geometry is of subtype~II(d) ${\Leftrightarrow\Psi_{2^{ij}} =0 \Leftrightarrow}$
\begin{equation}
F_{pq}\equiv f_{[p,q]}=0\,. \label{RTIId}
\end{equation}

\item
The type~III Robinson--Trautman geometry is of subtype~III(a) ${\Leftrightarrow{\Psi_{3T^i}=0} \Leftrightarrow}$
\begin{eqnarray}
\alpha_{,q}+f_{q}\,\alpha = 0 \,, \qquad \beta_{,q}+f_{q,u} = \mathcal{T}_q \,, \label{RTIIIa}
\end{eqnarray}
where
\begin{eqnarray}
\mathcal{T}_q \equiv {\textstyle 2\,e_{q}\big(\alpha-\frac{1}{4}f^p f_p\big)+\frac{1}{2}e^pf_pf_q-f^p E^{_\mathrm{RT}}_{pq}-\frac{2}{D-3}\,X_q} \,, \label{RTdefTq}
\end{eqnarray}
with ${X_q=h^{pm}X^{_\mathrm{RT}}_{pmq}=g^{pm}X_{pmq}\,}$,
\begin{equation}
\fl X^{_\mathrm{RT}}_{pmq}={\textstyle h_{p[m,u||q]}+e_{[q,m]||p}+e^{_\mathrm{RT}}_{p[m}f_{q]}-e_{[q}f_{m]||p}-\frac{1}{2}e_{[q}f_{m]}f_p} \,.
\end{equation}
\item
The type~III Robinson--Trautman geometry is of subtype~III(b) ${\Leftrightarrow{\tilde\Psi_{3^{ijk}}=0} \Leftrightarrow}$
\begin{equation}
{\textstyle X^{_\mathrm{RT}}_{pmq}=\frac{1}{D-3}\big( h_{pm}\,X_q-h_{pq}\,X_m\big)} \,. \label{RTIIIb}
\end{equation}

\item The type~N Robinson--Trautman geometry is described by the symmetric traceless matrix ${\Psi_{4^{ij}}}$, which is completely determined by
\begin{eqnarray}
\fl W_{pq} &=& \Big({\textstyle -\frac{1}{2}\gamma_{||p||q}+\frac{1}{2}\gamma_{,(p}f_{q)}+\frac{1}{2}\gamma f_{(p||q)}
+\frac{1}{2}\beta\,e^{_\mathrm{RT}}_{pq} +\big(\alpha-\frac{1}{4}f^nf_n\big)e_{p}e_{q}+e_{(p,u||q)}} \nonumber \\
\fl && \hspace{1.6mm} {\textstyle -\frac{1}{2}h_{pq,uu}+\frac{1}{4}e^{n}e_{n}\,f_{p}f_{q}-e^{n}E^{_\mathrm{RT}}_{n(p}f_{q)} +h^{mn}E^{_\mathrm{RT}}_{mp}E^{_\mathrm{RT}}_{nq}-\frac{2}{D-3}\,X_{(p}\,e_{q)}}\Big)\,r^2 \nonumber\\
\fl && {\textstyle +\frac{1}{2}}\big(\,2\alpha\, e^{_\mathrm{RT}}_{pq}-\mathcal{T}_{(p||q)}-\mathcal{T}_{(p}f_{q)}-f_{(p,u||q)}-f_{(p} f_{q),u}\big)\,r\,. \label{RTNWpq}
\end{eqnarray}

\item The type~N becomes type~O ${\Leftrightarrow{\Psi_{4^{ij}}=0}\Leftrightarrow W_{pq}=\frac{1}{D-2}\, h_{pq}\,h^{mn}W_{mn}}$, where $W_{pq}$ is given by (\ref{RTNWpq}).
\end{itemize}
\smallskip

\noindent
This completes the classification of \emph{principal} alignment (sub)types of the Weyl tensor of Robinson--Trautman geometries with expansion (\ref{Theta_je_1/r}) and the multiple WAND~${\mathbf{k}}$.

The \emph{secondary} alignment (sub)types with the additional WAND ${\boldl=\frac{1}{2}g_{uu}\mathbf{\partial}_r+\mathbf{\partial}_u}$ are obtained when the conditions summarized in table~\ref{SecClassSch} are satisfied, namely:
\smallskip
\begin{itemize}
\item
The type II Robinson--Trautman geometry is of type~$\mathrm{II}_i$ ${\Leftrightarrow \Psi_{4^{ij}}=0 \Leftrightarrow}$
\begin{equation}
W_{pq}={\textstyle\frac{1}{D-2}}\, h_{pq}\,h^{mn}W_{mn}\,, \label{RTIIi}
\end{equation}
where $W_{pq}$ is given by (\ref{WpqIIRT}).

\item
The type III Robinson--Trautman geometry is of type~$\mathrm{III}_i$ ${\Leftrightarrow \Psi_{4^{ij}}=0 \Leftrightarrow}$ the condition (\ref{RTIIi}) is satisfied, where $W_{pq}$ is given by
\begin{eqnarray}
\fl W_{pq}&=&\Big({\textstyle -\frac{1}{2}\gamma_{||p||q}+\frac{1}{2}\gamma_{,(p}f_{q)}+\frac{1}{2}\gamma f_{(p||q)}
+\frac{1}{2}\beta\,e^{_\mathrm{RT}}_{pq}+\beta_{,(p}e_{q)} -\big(\alpha-\frac{1}{4}f^nf_n\big)e_{p}e_{q}} \nonumber \\
\fl && \hspace{2.3mm} {\textstyle +e_{(p,u||q)}+e_{(p}f_{q),u}-\frac{1}{2}h_{pq,uu}+\frac{1}{4}e^{n}e_{n}\,f_{p}f_{q}-\frac{1}{2}e^nf_ne_{(p}f_{q)}} \nonumber \\
\fl && \hspace{2.3mm} {\textstyle +h^{mn}E^{_\mathrm{RT}}_{mp}E^{_\mathrm{RT}}_{nq}+f^nE^{_\mathrm{RT}}_{n(p}e_{q)}-e^nE^{_\mathrm{RT}}_{n(p}f_{q)}\Big)\,r^2} \nonumber\\
\fl && {\textstyle
-\big(\frac{1}{2}\beta_{||p||q}+\frac{1}{2}\beta_{,(p}f_{q)}
-\alpha\, e^{_\mathrm{RT}}_{pq}-2\alpha_{,(p}e_{q)}-2\alpha\, e_{(p}f_{q)}
+f_{(p,u||q)}+f_{(p} f_{q),u}\big)\,r} \nonumber\\
\fl && {\textstyle
-\frac{1}{2}\alpha_{||p||q}-\frac{1}{2}\alpha f_{(p||q)} -\frac{3}{2}\alpha_{,(p}f_{q)}-\alpha f_p f_q}\,. \label{RTIIIi}
\end{eqnarray}
This, in fact, is a general form of $W_{pq}$ for the type~III spacetimes.

\item
The type II Robinson--Trautman geometry is of type~$\mathrm{D}$ with respect to the double WAND~${\boldl=\frac{1}{2}g_{uu}\mathbf{\partial}_r+\mathbf{\partial}_u}$ ${\Leftrightarrow
\Psi_{3T^{i}}=\tilde{\Psi}_{3^{ijk}}=0}$ and ${\Psi_{4^{ij}}=0} \Leftrightarrow$
\begin{eqnarray}
&&\fl
{\textstyle X^{_\mathrm{RT}}_{pmq}=\frac{1}{D-3}\big( h_{pm}\,X^{_\mathrm{RT}}_q
-h_{pq}\,X^{_\mathrm{RT}}_m\big)}\,,\qquad
{\textstyle Y^{_\mathrm{RT}}_{pmq}=\frac{1}{D-3}\big( h_{pm}\,Y^{_\mathrm{RT}}_q
-h_{pq}\,Y^{_\mathrm{RT}}_m\big)}\,, \nonumber\\
&&\fl
V_p=0\,,\qquad W_{pq}={\textstyle\frac{1}{D-2}}\, h_{pq}\,h^{mn}W_{mn}\,, \label{RTD}
\end{eqnarray}
where the corresponding functions are given by (\ref{XpmqIIRTRT}), (\ref{YpmqIIRTRT}), (\ref{XYRTq}), (\ref{VpIIRT}) and (\ref{WpqIIRT}).
The particular subtypes D(a), D(b), D(c), D(d) and their various combinations occur if the additional conditions (\ref{RTIIa})--(\ref{RTIId}) are also valid.
\end{itemize}

\smallskip

\section{Example: vacuum Robinson--Trautman spacetimes}
\label{RTexample}

Motivated by our previous studies \cite{PodOrt06,SvarcPodolsky:2014} of Robinson--Trautman spacetimes in general relativity (extended to any dimension ${D\ge4}$), let us consider a metric of the form
\begin{equation}
\dd s^2 = r^2\,h_{pq}\, (\dd x^p+e^{p}\, \dd u)(\dd x^q+e^{q}\, \dd u) -2\,\dd u\dd r-g^{rr}\, \dd u^2 \,, \label{RTmetricFin}
\end{equation}
where
\begin{equation}
\fl g^{rr}\!=\frac{\mathcal{R}}{(D-2)(D-3)}+\frac{b(u)}{r^{D-3}}-\frac{2}{D-2}\big(e^{p}\,\!_{||p}-{\textstyle \frac{1}{2}}h^{pq}h_{pq,u}\big)\,r-\frac{2\Lambda}{(D-1)(D-2)}\,r^2.
\label{grrfinal}
\end{equation}
In fact, this is the most general Robinson--Trautman vacuum line element in Einstein's theory (extended to an arbitrary dimension $D$), with a cosmological constant $\Lambda$ and possibly a pure radiation field aligned with ${\mathbf{k}=\mathbf{\partial}_r}$.

Employing the results of the previous section, it is straightforward to obtain explicit conditions under which this geometry becomes a specific algebraic type. Since
\begin{equation}
f_p=0 \label{fpfina}
\end{equation}
and ${g_{up}= e_p(u,x)\,r^2}$, the condition (\ref{RTgup}) is satisfied, so that the spacetime is of Weyl type II (or more special) with respect to the multiple WAND~${\mathbf{k}}$. Moreover, ${g_{uu}=r^2\,e^ne_n-g^{rr}}$, that is
\begin{equation}\label{grrfina}
{\textstyle
g_{uu}=-a-b\,r^{3-D}-c\,r+\gamma\,r^2 }\,,
\end{equation}
where
\begin{eqnarray}
a &=& \frac{\mathcal{R}}{(D-2)(D-3)}\, \,, \label{aaa}\\
c &=& -\frac{2}{D-2}\big(e^{n}\,\!_{||n}-\frac{1}{2}h^{mn}h_{mn,u}\big) \,, \label{ccc}\\
\gamma &=& e^ne_n+\frac{2\Lambda}{(D-1)(D-2)}\, \,. \label{ac}
\end{eqnarray}
For ${f_p=0}$ and $g_{uu}$ of the form (\ref{grrfina}) we obtain from (\ref{PIIRT})--(\ref{WpqIIRT}) that
\begin{eqnarray}
\fl P &=& {-\textstyle \frac{1}{2}(D-1)(D-2)\,b\,r^{1-D}} \,, \label{PIIRT f0}\\
\fl Q_{pq} &=& {\textstyle \mathcal{R}_{pq}}\,, \label{QpqIIRT f0}\\
\fl F_{pq} &=& 0 \,, \label{FpqIIRT f0}\\
\fl V_p &=& {\textstyle \big(a\,e_p-\frac{1}{2} c_{,p}+\frac{1}{D-3}X_p\big)-a_{,p}\,r^{-1} +\frac{1}{2}(D-1)(D-2)\,b\,e_p\,r^{3-D}} \,, \label{VpIIRT f0}\\
\fl X_{pmq}\!&=& \big(h_{p[m,u||q]}+e_{[q,m]||p}\big)\,r^2 \,, \label{XpmqIIRT f0}\\
\fl W_{pq}&=& {\textstyle \frac{1}{2}a_{||p||q}+\big(\frac{1}{2}c_{||p||q}
-a\,e_{pq}^{_\mathrm{RT}}-2a_{,(p}e_{q)}\big)\,r} \nonumber \\
\fl && {\textstyle+\big(\!-\frac{1}{2}(e^ne_n)_{||p||q}-\frac{1}{2}c\,e_{pq}^{_\mathrm{RT}}
+ae_p e_q-c_{,(p}e_{q)}} \nonumber \\
\fl && {\textstyle\hspace{5.3mm}
+e_{(p,u||q)}-\frac{1}{2}h_{pq,uu}+h^{mn}E^{_\mathrm{RT}}_{mp}E^{_\mathrm{RT}}_{nq} \big)\,r^2}\nonumber \\
\fl && {\textstyle-\frac{1}{2}(D-1)\,b\,e_{pq}^{_\mathrm{RT}}\,r^{4-D}
+\frac{1}{2}(D-1)(D-2)\,b\,e_p e_q\,r^{5-D}} \,. \label{WpqIIRT f0}
\end{eqnarray}
In view of (\ref{Psi0ij})--(\ref{Psi4ij}), the metric (\ref{RTmetricFin}), (\ref{grrfinal}) is thus of
\begin{eqnarray}
\fl \quad\bullet\quad \hbox{subtype~II(a)} \quad &\Leftrightarrow& \quad \,b(u)=0\,, \label{RTIIaex}\\
\fl \quad\bullet\quad \hbox{subtype~II(b)} \quad &\Leftrightarrow& \quad \mathcal{R}_{pq}=\frac{h_{pq}}{D-2}\,\mathcal{R}\,, \label{RTIIbex}\\
\fl \quad\bullet\quad \hbox{subtype~II(c)} \quad &\Leftrightarrow& \quad \mathcal{C}_{mpnq}=0 \,, \label{RTIIcex}\\
\fl \quad\bullet\quad \hbox{subtype~II(d)} && \quad\,\,\hbox{always}\,. \label{RTIIdex}
\end{eqnarray}

Interestingly, it follows from the contraction of Bianchi identities and condition (\ref{RTIIbex}), which is identically satisfied in ${D=4}$, that
\begin{equation}
0=h^{mn}h^{pq}(\mathcal{R}_{mpnq||k}+\mathcal{R}_{mpqk||n}+\mathcal{R}_{mpkn||q})={\textstyle \frac{D-4}{D-2}}\,\mathcal{R}_{,k} \,.
\end{equation}
Thus, \emph{any} Robinson--Trautman ${D>4}$ geometry (\ref{RTmetricFin}), (\ref{grrfinal}) of algebraic subtype~II(b) must have ${\mathcal{R}_{,k}=0}$. Moreover
\begin{equation}
\mathcal{C}_{mpnq} = \mathcal{R}_{mpnq}-{\textstyle\frac{1}{(D-2)(D-3)}}\,\mathcal{R}\,(h_{mn}h_{pq}-h_{mq}h_{np}) \,. \label{CmpnqIIRT f0 EEq}
\end{equation}

The condition ${\mathcal{C}_{mpnq} = 0}$ for subtype~II(c) is always satisfied in ${D=4}$ and ${D=5}$.

We thus immediately infer that subtype II(bc)$\equiv$II(bcd) occurs if, and only if, the ${(D-2)}$-dimensional transverse space has a \emph{constant curvature}, that is
\begin{equation}
\mathcal{R}_{mpnq} = {\textstyle\frac{1}{(D-2)(D-3)}}\,\mathcal{R}(u)\,(h_{mn}h_{pq}-h_{mq}h_{np}) \,.
\label{constcurvature}
\end{equation}
In such a case the metric can be written as ${h_{pq}=P^{-2}\delta_{pq}}$, where ${P=1+\frac{1}{4}K\,\delta_{mn}x^m x^n}$, ${K=\frac{1}{(D-2)(D-3)}\,\mathcal{R}(u)}$, see \cite{PodOrt06,OrtPodZof08}.

\noindent
The conditions for type~III subtypes are then
\begin{eqnarray}
\fl \quad\bullet\quad \hbox{subtype~III(a)} \quad &\Leftrightarrow& \quad \mathcal{R}_{,q}=0 \quad\hbox{and}\quad
c_{,q}=2\,a\,e_q+\frac{2}{D-3}X_q\,, \label{RTIIIaex}\\
&& \hspace{4.5mm}\hbox{where}\quad X_q=h^{pm}X^{_\mathrm{RT}}_{pmq}\,,\quad X^{_\mathrm{RT}}_{pmq}=h_{p[m,u||q]}+e_{[q,m]||p}\,, \nonumber\\
\fl \quad\bullet\quad \hbox{subtype~III(b)} \quad &\Leftrightarrow& \quad { X^{_\mathrm{RT}}_{pmq}=\frac{1}{D-3}\big( h_{pm}\,X_q-h_{pq}\,X_m\big)}\,. \label{RTIIIbex}
\end{eqnarray}

\noindent
The type~$\mathrm{N}$ is obtained by applying \emph{all} the conditions (\ref{RTIIaex})--(\ref{RTIIdex}) and (\ref{RTIIIaex})--(\ref{RTIIIbex}). In such a case, the only non-trivial function (\ref{WpqIIRT f0}) reduces to
\begin{eqnarray}
W_{pq}&=& {\textstyle \big(\frac{1}{2}c_{||p||q}
-a\,e_{pq}^{_\mathrm{RT}}\big)\,r} \nonumber \\
\fl && {\textstyle+\Big(\!-\frac{1}{2}(e^ne_n)_{||p||q}-\frac{1}{2}c\,e_{pq}^{_\mathrm{RT}}
+a\,e_p e_q-c_{,(p}e_{q)}} \nonumber\\
&& {\textstyle\hspace{5.2mm}
+e_{(p,u||q)}-\frac{1}{2}h_{pq,uu}+h^{mn}E^{_\mathrm{RT}}_{mp}E^{_\mathrm{RT}}_{nq} \Big)\,r^2}\,. \label{WpqNRT f0}
\end{eqnarray}
Such a geometry is of type~$\mathrm{O}$ when ${W_{pq}=\frac{1}{D-2}\, h_{pq}\,h^{mn}W_{mn}}$.

The secondary alignment types~$\mathrm{II}_i$, $\mathrm{III}_i$, $\mathrm{D}$ arise when the conditions (\ref{RTIIi}), (\ref{RTIIIi}), (\ref{RTD}) are satisfied, respectively, in which the key functions are given by (\ref{VpIIRT f0})--(\ref{WpqIIRT f0}).

We thus conclude that the Robinson--Trautman geometry of the form (\ref{RTmetricFin}), (\ref{grrfinal}) generally admits \emph{all} the above mentioned algebraic types and subtypes. Of course, specific \emph{field equations impose additional constrains} that may exclude some of the (sub)types. To illustrate this effect, let us now restrict ourselves to the most important case, namely vacuum spacetimes in the \emph{Einstein theory}.

\subsection{Most general Robinson--Trautman vacuum spacetimes}
\label{EinsteinRTexample}

As shown in \cite{SvarcPodolsky:2014}, a fully general  Robinson--Trautman vacuum solution in the Einstein theory (including $\Lambda$) is given by (\ref{RTmetricFin}), (\ref{grrfinal}) where the metric functions are restricted by the constraints
\begin{eqnarray}
&& \mathcal{R}_{pq}={\textstyle \frac{1}{D-2}}\,h_{pq}\,\mathcal{R}\,, \label{EEq Rpq} \\
&& h_{pq,u}=2\,e_{(p||q)}+c\,h_{pq}\,, \label{hpqu} \\
&& (D-4)\,\mathcal{R}_{,p}=0 \,, \label{Rp0} \\
&& {\textstyle h^{mn}\,a_{||m||n}+\frac{1}{2}(D-1)(D-2)\,b\,c+(D-2)\,b_{,u}=0 } \,, \label{RTEq} \end{eqnarray}
with $a$ and $c$ defined in (\ref{aaa}) and (\ref{ccc}).

\newpage

Using conditions (\ref{EEq Rpq})--(\ref{RTEq}) and
\begin{eqnarray}
&& e^{_\mathrm{RT}}_{pq}= -{\textstyle\frac{1}{2}}c\,h_{pq}\,,\quad
E^{_\mathrm{RT}}_{pq} = e_{p||q}+{\textstyle\frac{1}{2}}c\,h_{pq}\,,\\
&& h_{pq,uu}=2c\,e_{(p||q)}+2e_{(p||q),u}+(c^2+c_{,u})\,h_{pq} \,, \\
&& X^{_\mathrm{RT}}_{pmq} = e^{n}\mathcal{R}_{npmq}+c_{,[q}h_{m]p}\,,\quad
X_q = {\textstyle (D-3)(-a\,e_q+\frac{1}{2}\,c_{,q})} \,, \label{ecka EEq}
\end{eqnarray}
functions (\ref{VpIIRT f0})--(\ref{WpqIIRT f0}) which, together with
(\ref{PIIRT f0})--(\ref{FpqIIRT f0}), characterize the algebraic structure of the Weyl tensor become
\begin{eqnarray}
\fl V_p &=& {\textstyle -a_{,p}\,r^{-1} +\frac{1}{2}(D-1)(D-2)\,b\,e_p\,r^{3-D}} \,, \label{VpIIRT f0 EEq}\\
\fl X_{pmq}\!&=& \big(e^n\,\mathcal{R}_{npmq}+c_{,[q}h_{m]p}\big)\,r^2 \,, \label{XpmqIIRT f0 EEq}\\
\fl W_{pq}&=& {\textstyle \frac{1}{2}a_{||p||q}+\big(\frac{1}{2}c_{||p||q}-2a_{,(p}e_{q)}\big)\,r} \nonumber \\
\fl && {\textstyle +e^me^n\,\mathcal{C}_{mpnq}\,r^2+\frac{1}{2}(D-1)(D-2)\,b\,e_pe_q\,r^{5-D}} \nonumber \\
\fl && {\textstyle + h_{pq}\,\big[ \frac{1}{2}a\,c\,r+\big(a\,e^ne_n-\frac{1}{2}e^nc_{,n}-\frac{1}{2}c_{,u}\big)\,r^2
+\frac{1}{4}(D-1)\,b\,c\,r^{4-D}\big]}\,. \label{WpqIIRT f0 EEq}
\end{eqnarray}
To derive the last expression we have used the fact that
${e_{(p,u||q)}-e_{(p||q),u}=e_m\,T^m_{pq}\,}$, where the tensor ${T^m_{pq}\equiv \!\,^{S}\Gamma^m_{pq,u}}$ can be written as
\begin{equation}
T^m_{pq}=h^{mn}\big[e_{n||(p||q)}-e^k \,\mathcal{R}_{k(pq)n}\big]+\delta^m_{\ (p}\,c_{,q)}-{\textstyle \frac{1}{2}} h^{mn}h_{pq}\,c_{,n} \,, \label{Ttens}
\end{equation}
see Appendix~A of \cite{SvarcPodolsky:2014}. The non-vanishing Weyl tensor components with respect to the frame (\ref{nat null frame}), sorted by the boost weight, are thus
\begin{eqnarray}
\fl \Psi_{2S} &=& {\textstyle -\frac{1}{2}(D-2)(D-3)\,b\,r^{1-D}} \,, \\
\fl \tilde{\Psi}_{2^{ijkl}} &=& m_i^m m_j^p m_k^n m_l^q\,\mathcal{C}_{mpnq}\,r^2 \,, \\
\fl \Psi_{3T^i} &=& {\textstyle m_i^p\,\big[\frac{1}{2}(D-1)(D-3)\,b\,e_p\,r^{3-D}-\frac{D-3}{D-2}\,a_{,p}\,r^{-1}\big]} \,, \\
\fl \tilde{\Psi}_{3^{ijk}} &=& {\textstyle m_i^pm_j^mm_k^q\,e^n\,\mathcal{C}_{npmq}\,r^2} \,, \\
\fl \Psi_{4^{ij}} &=& {\textstyle m_i^pm_j^q\Big[\,\frac{1}{2}\big(a_{||p||q}-\frac{1}{D-2}\,h_{pq}\,h^{mn}a_{||m||n}\big)
+e^me^n\, \mathcal{C}_{mpnq}\,r^2 } \nonumber\\
\fl &&\hspace{8.5mm}{\textstyle +\big[\frac{1}{2}\big(c_{||p||q}-\frac{1}{D-2}\,h_{pq}\,h^{mn}c_{||m||n}\big)-2\big(a_{,(p}e_{q)}-\frac{1}{D-2}\,h_{pq}\,e^n\,a_{,n}\big)\big]\,r} \nonumber \\
\fl &&\hspace{8.5mm}{\textstyle+\frac{1}{2}(D-1)(D-2)\,b\,\big(e_pe_q-\frac{1}{D-2}\,h_{pq}\,e^ne_n\big)\,r^{5-D} \Big]} \,.
\end{eqnarray}

It is now convenient to perform a null rotation of the frame (\ref{nat null frame}) with the privileged null vector $\boldk$ fixed,
${\boldl^{\prime} = \boldl + \sqrt2\, L^i \boldm_{i} + |L|^2\, \boldk}$, ${\boldm^{\prime}_{i} = \boldm_{i} + \sqrt2\, L_i\,\boldk}$, see (C1) in Appendix~C of \cite{PodolskySvarc:2012},
and the parameters ${L_i\equiv-\frac{1}{\sqrt{2}}\,r^2\,e_{p}\,m_{i}^{p}}$. The \emph{new} null frame is
\begin{equation}
\boldk^{\prime}=\mathbf{\partial}_r \,, \ \quad \boldl^{\prime}=-{\textstyle \frac{1}{2}}g^{rr}\,\mathbf{\partial}_r+\mathbf{\partial}_u-e^{p}\,\mathbf{\partial}_p \,, \ \quad \boldm^{\prime}_i=m_i^p\,\mathbf{\partial}_p \,, \label{null frame - simple m}
\end{equation}
and using the expressions (C5) in \cite{PodolskySvarc:2012} the nonvanishing irreducible Weyl scalars in such a frame simplify considerably to
\begin{eqnarray}
\fl \Psi_{2S}^{\prime} &=& {\textstyle -\frac{1}{2}(D-2)(D-3)\,b\,r^{1-D}} \,, \label{Weyl_scalars_simple1}\\
\fl \tilde{\Psi}_{2^{ijkl}}^{\prime} &=& m_i^m m_j^p m_k^n m_l^q\,\mathcal{C}_{mpnq}\,r^2 \,,
\label{Weyl_scalars_simple2}\\
\fl \Psi_{3T^i}^{\prime} &=& {\textstyle -m_i^p\,\frac{D-3}{D-2}\,a_{,p}\,r^{-1}} \,,
\label{Weyl_scalars_simple3}\\
\fl \Psi_{4^{ij}}^{\prime} &=& {\textstyle \frac{1}{2}m_i^pm_j^q\,\Big[\big(a_{||p||q}+c_{||p||q}\,r\big)-\frac{1}{D-2}\,h_{pq}\,h^{mn}\big(a_{||m||n}+c_{||m||n}\,r\big)\Big]} \,, \label{Weyl_scalars_simple4}
\end{eqnarray}
with ${\tilde{\Psi}_{3^{ijk}}^{\prime}=0}$. Notice, interestingly, that the last term can be rewritten as
\begin{equation}
\Psi_{4^{ij}}^{\prime} = {\textstyle \frac{1}{2}m_i^pm_j^q\,\Big[\,g^{rr}_{\hspace{2.4mm}||p||q}-\frac{1}{D-2}\,h_{pq}\,h^{mn}\,g^{rr}_{\hspace{2.4mm}||m||n}\Big]} \,.
\end{equation}
The gravitational wave amplitude matrix ${\Psi_{4^{ij}}^{\prime}}$ (which is symmetric and traceless) is thus directly determined by \emph{the second spatial derivatives of the contravariant metric coefficient} ${g^{rr}}$, see (\ref{grrfinal}).
\newpage

To prove the \emph{non-existence of} type~N \emph{and} type~II \emph{vacuum solutions in} ${D>4}$ it is now crucial to prove the identity
\begin{equation}
{\textstyle (D-4)\,\big(c_{||p||q}-\frac{1}{D-2}\,h_{pq}\,h^{mn}c_{||m||n}\big)=0} \,. \label{cpq id}
\end{equation}
This follows from the $u$-derivative of the condition (\ref{EEq Rpq}), namely
\begin{equation}
\fl
{\textstyle \mathcal{R}_{pq,u}=\frac{1}{D-2}\big(\,h_{pq,u}\mathcal{R}+\frac{1}{D-2}\,h_{pq}\,h_{mn}\,{h^{mn}}_{,u}\,\mathcal{R}+h_{pq}\,h^{mn}\,\mathcal{R}_{mn,u}\big)} \,,
\end{equation}
which using ${{h^{mn}}_{,u}=-h^{mp}h^{nq}h_{pq,u}}$ and the constraint (\ref{hpqu}) can be rewritten as
\begin{equation}
\fl
{\textstyle \mathcal{R}_{pq,u}=\frac{1}{D-2}\big(2\,e_{(p||q)}\mathcal{R}-\frac{2}{D-2}\,h_{pq}\,{e^n}_{||n}\mathcal{R}+h_{pq}\,h^{mn}\,\mathcal{R}_{mn,u}\big)} \,. \label{Rpqu expl RHS}
\end{equation}
It remains to evaluate ${\mathcal{R}_{pq,u}}$. From the definition of the Ricci tensor it follows that
\begin{equation}
\fl
{\textstyle \mathcal{R}_{pq,u}=T^m_{pq||m}-T^m_{pm||q}} \,, \label{Rpqu}
\end{equation}
where ${T^m_{pq}\equiv\,^{S}\Gamma^m_{pq,u}}$ is a \emph{tensor} symmetric in ${p,q}$ and given by (\ref{Ttens}).
Using common relations for commutators of covariant derivatives and contracted Bianchi identities, see (3.2.3), (3.2.21), (3.2.16) in \cite{Wald:1984}, we obtain the derivatives of (\ref{Ttens})
\begin{equation}
\fl {\textstyle T^m_{pq||m}={e^n}_{||n||p||q}+\frac{2}{D-2}\,e_{(p||q)}\,\mathcal{R}
+\frac{1}{D-2}\,h_{pq}\,e^n\,\mathcal{R}_{,n}+c_{||p||q}-\frac{1}{2}h_{pq}\,h^{mn}c_{||m||n} } \,,
\end{equation}
\begin{equation}
\fl {\textstyle T^m_{pm||q}={e^n}_{||n||p||q}+\frac{1}{2}(D-2)\,c_{||p||q} } \,.
\end{equation}
Substituting into (\ref{Rpqu}), the $u$-derivative of the Ricci tensor becomes
\begin{equation}
\fl {\textstyle \mathcal{R}_{pq,u}=\frac{2}{D-2}\,e_{(p||q)}\,\mathcal{R}+\frac{1}{D-2}\,h_{pq}\,e^n\,\mathcal{R}_{,n}
-\frac{1}{2}(D-4)\,c_{||p||q}-\frac{1}{2}h_{pq}\,h^{mn}c_{||m||n}} \,, \label{Rpqu expl}
\end{equation}
and its trace reads
${h^{mn}\mathcal{R}_{mn,u}=\frac{2}{D-2}\,{e^n}_{||n}\,\mathcal{R}
+\,e^n\,\mathcal{R}_{,n}-(D-3)\,h^{mn}c_{||m||n}}$.
By~inserting these two expressions into (\ref{Rpqu expl RHS}), the identity (\ref{cpq id}) is proven.

Using (\ref{Weyl_scalars_simple1})--(\ref{Weyl_scalars_simple4}) with (\ref{Rp0}) and (\ref{cpq id}) it is now easy to determine explicitly the algebraic structure of all vacuum Robinson--Trautman spacetimes in the Einstein theory in any dimension $D$. The results are summarized in table~\ref{RT EEq summary}.
\begin{table}[b]
\begin{tabular}{c||l|ll}
{\small type}    & \multicolumn{1}{c|}{${D=4}$} & \multicolumn{2}{c}{${D>4}$}  \\ \hline\hline\hline
$\mathrm{II(a)}$ & ${b=0}$ & ${b=0}$ & $\Leftrightarrow$ \ $\mathrm{D(a)}$ \\ \hline
$\mathrm{II(b)}$ & {\small always} & {\small always} & $\Leftrightarrow$ \ $\mathrm{D(b)}$\\ \hline
$\mathrm{II(c)}$ & {\small always} & ${\mathcal{C}_{mpnq}=0}$ & $\Leftrightarrow$ \ $\mathrm{D(c)}$ \\ \hline
$\mathrm{II(d)}$ & {\small always} & {\small always} & $\Leftrightarrow$ \ $\mathrm{D(d)}$\\ \hline\hline\hline
$\mathrm{III}$ & \multicolumn{3}{c}{$\mathrm{II(abcd)}$} \\ \hline\hline
$\mathrm{III(a)}$ & ${\mathcal{R}_{,p}=0}$ & {\small always} & \\ \hline
$\mathrm{III(b)}$ & {\small always} & {\small always} & \\ \hline\hline\hline
$\mathrm{N}$ & \multicolumn{3}{c}{$\mathrm{III(ab)}$} \\ \hline\hline
$\mathrm{O}$ & ${c_{||p||q}=\frac{1}{D-2}\,h_{pq}\,h^{mn}c_{||m||n}}$ & {\small always} & $\Leftrightarrow$ \ $\mathrm{D(ac)}$ \\ \hline\hline\hline
$\mathrm{D}$ & ${\mathcal{R}_{,p}=0}$ \  {\small and} \  ${c_{||p||q}=\frac{1}{D-2}\,h_{pq}\,h^{mn}c_{||m||n}}$ & {\small always} &  \\
\end{tabular}
\caption{\label{RT EEq summary} The necessary and sufficient conditions for all possible algebraic (sub)types of the Robinson--Trautman vacuum solutions of Einstein's field equations. Some of them are always satisfied. The admissible algebraic structures of the Weyl tensor differ significantly in the case ${D=4}$ and in higher dimensions ${D>4}$.}
\end{table}

In ${D>4}$ it follows from (\ref{Rp0}) that ${a_{,p}=0}$, which, together with (\ref{cpq id}), implies ${\Psi_{3T^i}^{\prime} = 0 =\Psi_{4^{ij}}^{\prime} }$. This proves that \emph{there are no} type~N, type~III and type~II spacetimes in the Robinson--Trautman family in higher dimensions when the Einstein vacuum field equations are applied. Such a result is in full agreement with observations made in \cite{PodOrt06}.

Thus, \emph{all Robinson--Trautman vacuum solutions in ${D>4}$ are of type}~D (or type~O). The double degenerate WANDs are ${\boldk^{\prime}=\mathbf{k}=\mathbf{\partial}_r}$, ${\boldl^{\prime}=-\frac{1}{2}g^{rr}\,\mathbf{\partial}_r+\mathbf{\partial}_u-e^{p}\,\mathbf{\partial}_p\,}$, see (\ref{null frame - simple m}). It is also straightforward to identify the possible subtypes, namely D(a) and D(c) and O$\equiv$D(ac). In fact, the only non-vanishing Weyl scalars are
\begin{eqnarray}
\Psi_{2S}^{\prime} &=& -{\textstyle\frac{1}{2}}(D-2)(D-3)\,b\,r^{1-D} \,,\label{PsiRTDaex} \\
\tilde{\Psi}_{2^{ijkl}}^{\prime} &=& m_i^m m_j^p m_k^n m_l^q\,\,\mathcal{C}_{mpnq}\,r^2 \,, \label{PsiRTDcex}
\end{eqnarray}
so that all such spacetimes are of the subtype~D(bd). Clearly, there are only two algebraically distinct cases possible, namely
\smallskip
\begin{eqnarray}
\fl \quad\bullet\quad \hbox{subtype~D(a)$\equiv$D(abd)} \quad &\Leftrightarrow& \quad \,b=0\,, \label{RTDaex}\\
\fl \quad\bullet\quad \hbox{subtype~D(c)$\equiv$D(bcd)} \quad &\Leftrightarrow& \quad \mathcal{C}_{mpnq}=0 \,. \label{RTDcex}
\end{eqnarray}
The latter case (which necessarily occurs in dimension ${D=5}$) admits just the scalar
${\Psi_{2S}^{\prime}}$ given by (\ref{PsiRTDaex}). Moreover, in view of (\ref{CmpnqIIRT f0 EEq}) relation (\ref{constcurvature}) must hold which means that the \emph{transverse Riemannian space has a constant curvature}. Such a family of Robinson--Trautman vacuum spacetimes contains generalizations of the Schwarzschild black hole of mass proportional to $b$. When both conditions (\ref{RTDaex}) and (\ref{RTDcex}) are satisfied, the corresponding spacetime is of type~O.

This generalizes, confirms and refines the conclusions of a previous work \cite{PodOrt06} where the Robinson--Trautman vacuum solutions with ${e_p=0}$ were studied, i.e., assuming the metric functions~$e_p$ can be \emph{globally removed}. Relation between the respective notations is
${h_{pq}=P^{-2}(u,x)\,\gamma_{pq}(x)}$ with ${\hbox{det}\,\gamma_{pq}=1}$, ${\,b(u)=-\mu(u)\,}$ and
${c=-2\,(\log P)_{,u}}$. It follows that the exceptional cases discussed in \cite{PodOrt06} with the functions $\mu$ and/or $\mathcal{C}_{mpnq}$ vanishing are, in fact, \emph{algebraically distinct subtypes}.

\section{Concluding summary}

We investigated the algebraic structure of a fully general class on non-twisting and shear-free geometries in an arbitrary dimension $D$, that is, the complete Robinson--Trautman and Kundt family. In particular:

\begin{itemize}

\item Using the Christoffel symbols we derived all coordinate components of the Riemann, Ricci and Weyl curvature tensors in an explicit form. These are presented in~\ref{appendixA}.

\item By projecting the Weyl tensor onto the natural null frame we evaluated the corresponding scalars of all boost weights. In contrast to a complicated form of the coordinate components, such Weyl scalars are, due to cancelation of many terms, surprisingly simple, see equations (\ref{Psi0ij})--(\ref{Psi4ij}) with (\ref{P})--(\ref{Wpq}).

\item Weyl scalars obtained in this manner directly determine the algebraic structure of the metric (\ref{general nontwist}) with (\ref{IntShearFreeCond}). Distinct algebraic types and subtypes are defined by the vanishing of these scalars (and their combinations), see tables~\ref{PriClassSch} and~\ref{SecClassSch}.

\item We proved that all non-twisting shear-free geometries are of type~I(b), or more special, with the WAND aligned along the optically privileged null direction ${\mathbf{k}}$.

\item We were able to explicitly derive the necessary and sufficient conditions of all principal alignment types such that the optically privileged null direction ${\mathbf{k}}$ is a multiple WAND. These algebraically special (sub)types are II, II(a), II(b), II(c), II(d), their combinations, III, III(a), III(b), N and O. See the explicit conditions given in section~\ref{multipleWAND} and table~\ref{AlgSpecClassSch}.

\item In section~\ref{otherWANDl} we also identified the secondary alignment types for which there exists an additional specific WAND~$\boldl$ distinct from the (multiple) WAND $\mathbf{k}$, namely $\mathrm{I}_i$, $\mathrm{II}_i$, $\mathrm{III}_i$ or~D with a double~$\boldl$. Moreover, there are various subtypes, namely  $\mathrm{II(a)}_i$, $\mathrm{II(b)}_i$, $\mathrm{II(c)}_i$, $\mathrm{II(d)}_i$ (or their combinations), $\mathrm{III(a)}_i$, $\mathrm{III(b)}_i$ and D(a), D(b), D(c), D(d).

\item The Kundt family, which is the nonexpanding (${\Theta=0}$) subclass of the non-twisting shear-free geometries, is studied in section \ref{Kundt_geometries}. The corresponding conditions for algebraic types and subtypes are simplified, and they fully agree with those obtained previously in \cite{PodolskySvarc:2013a}.

\item The algebraic structure of the general Robinson--Trautman class with an arbitrary expansion scalar ${\Theta\not=0}$ is described in section~\ref{multipleWAND}. The special case ${\Theta=1/r}$ is investigated in section~\ref{RT_geometries}. In fact, this is an important subcase, as such Robinson--Trautman geometries are of Riemann type~I and also of Ricci type~I.

\item No field equations have been employed in these calculations and discussions. All  results are thus ``purely geometrical'', i.e., they can be applied in any metric theory of gravity that admits non-twisting and shear-free geometries.

\item Of course, there are specific constraints on admissible algebraic types imposed by the field equations. To illustrate this, in section~\ref{RTexample} we investigated an important example, namely the Robinson--Trautman vacuum solutions in Einstein's theory. We proved that in all dimensions higher than four there exist only types D(a)$\equiv$D(abd), D(c)$\equiv$D(bcd) and O of such spacetimes. This is in striking contrast to the classical ${D=4}$ case, which is much richer, see table~\ref{RT EEq summary}.

\end{itemize}

\section*{Acknowledgements}
J.P. has been supported by the grant GA\v{C}R P203/12/0118, R.\v{S}. by the project UNCE~204020/2012 and the Czech--Austrian MOBILITY grant 7AMB13AT003.

\appendix

\section{Explicit curvature tensors for a general non-twisting and shear-free geometry}
\label{appendixA}

After applying the shear-free condition (\ref{shearfree condition}) the Christoffel symbols for the general non-twisting geometry (\ref{general nontwist}) are
\begin{eqnarray}
&& {\textstyle \Gamma^r_{rr} = 0} \,, \label{ChristoffelBegin} \\
&& {\textstyle \Gamma^r_{ru} = -\frac{1}{2}g_{uu,r}+\frac{1}{2}g^{rn}g_{un,r}} \,, \\
&& {\textstyle \Gamma^r_{rp} = -\frac{1}{2}g_{up,r}+\Theta g_{up}} \,, \\
&& {\textstyle \Gamma^r_{uu} = \frac{1}{2}\big[-g^{rr}g_{uu,r}-g_{uu,u}+g^{rn}(2g_{un,u}-g_{uu,n})\big]} \,, \\
&& {\textstyle \Gamma^r_{up} = \frac{1}{2}\big[-g^{rr}g_{up,r}-g_{uu,p}+g^{rn}(2g_{u[n,p]}+g_{np,u})\big]} \,, \\
&& {\textstyle \Gamma^r_{pq} = -\Theta g^{rr}g_{pq}-g_{u(p||q)}+\frac{1}{2}g_{pq,u}} \,, \\
&& {\textstyle \Gamma^u_{rr}=\Gamma^u_{ru}=\Gamma^u_{rp} = 0} \,, \\
&& {\textstyle \Gamma^u_{uu} = \frac{1}{2}g_{uu,r}} \,, \\
&& {\textstyle \Gamma^u_{up} = \frac{1}{2}g_{up,r}} \,, \\
&& {\textstyle \Gamma^u_{pq} = \Theta g_{pq}} \,, \\
&& {\textstyle \Gamma^m_{rr} = 0} \,, \\
&& {\textstyle \Gamma^m_{ru} = \frac{1}{2}g^{mn}g_{un,r}} \,, \\
&& {\textstyle \Gamma^m_{rp} = \Theta\delta^m_p} \,, \\
&& {\textstyle \Gamma^m_{uu} = \frac{1}{2}\big[-g^{rm}g_{uu,r}+g^{mn}(2g_{un,u}-g_{uu,n})\big]} \,, \\
&& {\textstyle \Gamma^m_{up} = \frac{1}{2}\big[-g^{rm}g_{up,r}+g^{mn}(2g_{u[n,p]}+g_{np,u})\big]} \,, \\
&& {\textstyle \Gamma^m_{pq} = -\Theta g^{rm}g_{pq}+\,^{S}\Gamma^m_{pq}} \,. \label{ChristoffelEnd}
\end{eqnarray}
The Riemann curvature tensor components then read
\begin{eqnarray}
&& \fl R_{rprq} = {\textstyle -\big(\Theta_{,r}+\Theta^2\big)g_{pq}} \,, \label{Riem rprq}\\
&& \fl R_{rpru} = {\textstyle -\frac{1}{2}g_{up,rr}+\frac{1}{2}\Theta g_{up,r}} \,, \\
&& \fl R_{rpmq} = {\textstyle 2g_{p[m}\Theta_{,q]}-2\Theta^2g_{p[m}g_{q]u}+\Theta g_{p[m}g_{q]u,r}} \,, \\
&& \fl R_{ruru} = {\textstyle -\frac{1}{2}g_{uu,rr}+\frac{1}{4}g^{mn}g_{um,r}g_{un,r}} \,, \\
&& \fl R_{rpuq} = {\textstyle \frac{1}{2}g_{up,r||q}+\frac{1}{4}g_{up,r}g_{uq,r}-g_{pq}\Theta_{,u}} \nonumber \\
&& \fl \hspace{13.0mm} {\textstyle -\frac{1}{2}\Theta\big(g_{pq,u}+g_{pq}g_{uu,r}+g_{uq}g_{up,r}-g_{pq}g^{rn}g_{un,r}+2g_{u[p,q]}\big)} \,, \\
&& \fl R_{rupq} = {\textstyle g_{u[p,q],r}+\Theta\big(g_{u[p}g_{q]u,r}-2g_{u[p,q]}\big)} \,, \\
&& \fl R_{mpnq} = {\textstyle \,^{S}R_{mpnq}-\Theta^2g^{rr}(g_{mn}g_{pq}-g_{mq}g_{pn})} \nonumber \\
&& \fl \hspace{15.0mm} {\textstyle -\Theta\big(g_{mn}e_{pq}+g_{pq}e_{mn}-g_{mq}e_{pn}-g_{pn}e_{mq}\big)} \,, \\
&& \fl R_{ruup} = {\textstyle \frac{1}{2}(g_{uu,rp}-g_{up,ru})+\frac{1}{4}g^{rn}g_{un,r}g_{up,r}-\frac{1}{2}g^{mn}g_{um,r}E_{np}} \nonumber \\
&& \fl \hspace{15.0mm} {\textstyle +\Theta\big(g_{up,u}-\frac{1}{2}g_{uu,p}-\frac{1}{2}g_{up}g_{uu,r}\big)} \,, \\
&& \fl R_{upmq} = {\textstyle g_{p[m,u||q]}+g_{u[q,m]||p}+e_{p[m}g_{q]u,r}} \nonumber \\
&& \fl \hspace{15.0mm} {\textstyle +\Theta\big(g^{rr}g_{p[m}g_{q]u,r}+g_{uu,[q}g_{m]p}-2g^{rn}E_{n[q}g_{m]p}\big)} \,, \\
&& \fl R_{upuq} = {\textstyle -\frac{1}{2}(g_{uu})_{||p||q}+g_{u(p,u||q)}-\frac{1}{2}g_{pq,uu}+\frac{1}{4}g^{rr}g_{up,r}g_{uq,r}} \nonumber \\
&& \fl \hspace{13.7mm} {\textstyle -\frac{1}{2}g_{uu,r}e_{pq}+\frac{1}{2}g_{uu,(p}g_{q)u,r}-g^{rn}E_{n(p}g_{q)u,r}+g^{mn}E_{mp}E_{nq}} \nonumber \\
&& \fl \hspace{13.7mm} {\textstyle -\frac{1}{2}\Theta g_{pq}\big[g^{rr}g_{uu,r}+g_{uu,u}-g^{rn}(2g_{un,u}-g_{uu,n})\big]} \,.\label{Riem upuq}
\end{eqnarray}
The components of the Ricci tensor are
\begin{eqnarray}
&& \fl R_{rr} = {\textstyle -(D-2)\big(\Theta_{,r}+\Theta^2\big)} \,, \label{Ricci rr}\\
&& \fl R_{rp} = {\textstyle -\frac{1}{2}g_{up,rr}+g_{up}\Theta_{,r}-(D-3)\Theta_{,p}
+(D-2)\Theta^2g_{up}-\frac{1}{2}(D-4)\Theta g_{up,r}} \,, \label{Ricci rp} \\
&& \fl R_{ru} = {\textstyle -\frac{1}{2}g_{uu,rr}+\frac{1}{2}g^{rn}g_{un,rr}+\frac{1}{2}g^{mn}\big(g_{um,r||n}+g_{um,r}g_{un,r}\big)} \nonumber \\
&& \fl \hspace{10.5mm} {\textstyle -(D-2)\Theta_{,u}-\frac{1}{2}\Theta\big[g^{mn}g_{mn,u}-(D-4)g^{rn}g_{un,r}+(D-2)g_{uu,r}\big]} \,, \label{Ricci ru} \\
&& \fl R_{pq} = {\textstyle \,^{S}R_{pq}-f_{pq}-g_{pq}\big(g^{rr}\Theta_{,r}-2\Theta_{,u}+2g^{rn}\Theta_{,n}\big)+2g_{u(p}\Theta_{,q)}} \nonumber \\
&& \fl \hspace{10.5mm} {\textstyle +\Theta^2\big[2g_{pq}g^{rn}g_{un}-(D-2)g_{pq}g^{rr}-2g_{up}g_{uq}\big]} \nonumber \\
&& \fl \hspace{10.5mm} {\textstyle +\Theta\big[2g_{u(p||q)}+2g_{u(p}g_{q)u,r}-(D-2)e_{pq}} \nonumber \\
&& \fl \hspace{22.0mm} {\textstyle +g_{pq}\big(g_{uu,r}-2g^{rn}g_{un,r}-g^{mn}e_{mn}\big)\big]} \,, \label{Ricci pq} \\
&& \fl R_{up} = {\textstyle -\frac{1}{2}g^{rr}g_{up,rr}-\frac{1}{2}g_{uu,rp}+\frac{1}{2}g_{up,ru}+g^{rn}g_{u[n,p],r}-\frac{1}{2}g^{rn}(g_{up,r||n}+g_{un,r}g_{up,r})} \nonumber \\
&& \fl \hspace{10.5mm} {\textstyle +g^{mn}\big(\frac{1}{2}g_{um,r}g_{un||p}+g_{m[p,u||n]}+g_{u[m,p]||n}-\frac{1}{2}e_{mn}g_{up,r}\big)+g_{up}\Theta_{,u}}\nonumber \\
&& \fl \hspace{10.5mm} {\textstyle +\Theta\big[g_{up}g_{uu,r}+\frac{1}{2}(D-4)(g_{uu}g_{up,r}-g_{uu,p})-g_{up,u}-g^{rn}g_{un,r}g_{up}} \nonumber \\
&& \fl \hspace{10.5mm} {\textstyle +(D-6)g^{rn}(g_{u[n,p]}-\frac{1}{2}g_{un}g_{up,r})+\frac{1}{2}(D-2)g^{rn}g_{np,u}\big]}\,, \label{Ricci up} \\
&& \fl R_{uu} = {\textstyle -\frac{1}{2}g^{rr}g_{uu,rr}-g^{rn}g_{uu,rn}-\frac{1}{2}g^{mn}e_{mn}g_{uu,r}+g^{rn}g_{un,ru}-\frac{1}{2}g^{mn}g_{mn,uu}} \nonumber \\
&& \fl \hspace{10.5mm} {\textstyle +g^{mn}(g_{um,u||n}-\frac{1}{2}g_{uu||m||n})+\frac{1}{2}(g^{rr}g^{mn}-g^{rm}g^{rn})g_{um,r}g_{un,r}} \nonumber \\
&& \fl \hspace{10.5mm} {\textstyle +2g^{mn}g^{rp}g_{um,r}g_{u[n,p]}+\frac{1}{2}g^{mn}g_{um,r}g_{uu,n}+g^{mn}g^{pq}E_{pm}E_{qn}} \nonumber \\
&& \fl \hspace{10.5mm} {\textstyle +\frac{1}{2}\Theta\big[(D-4)g^{rn}(2g_{un,u}-g_{uu,n}-g_{un}g_{uu,r})} \nonumber \\
&& \fl \hspace{22.0mm} {\textstyle +(D-2)(g_{uu}g_{uu,r}-g_{uu,u})\big]} \,, \label{Ricci uu}
\end{eqnarray}
and the Ricci scalar is
\begin{eqnarray}
&& \fl R = {\textstyle \,^{S}R+g_{uu,rr}-2g^{rn}g_{un,rr}-2g^{mn}g_{um,r||n}-\frac{3}{2}g^{mn}g_{um,r}g_{un,r}} \nonumber \\
&& \fl \hspace{8.0mm} {\textstyle +2\Theta_{,r}\big[(D-2)g_{uu}-(D-3)g^{rn}g_{un}\big]+4(D-2)\Theta_{,u}-4(D-3)g^{rn}\Theta_{,n}} \nonumber \\
&& \fl \hspace{8.0mm} {\textstyle -\Theta^2\big[(D-1)(D-2)g^{rr}-2(2D-5)g^{rn}g_{un}\big]} \nonumber \\
&& \fl \hspace{8.0mm} {\textstyle +\Theta\big[2(D-2)g_{uu,r}-2(2D-7)g^{rn}g_{un,r}} \nonumber \\
&& \fl \hspace{16.0mm} {\textstyle +(D-1)g^{mn}g_{mn,u}-2(D-3)g^{mn}g_{um||n}\big]} \,.
\end{eqnarray}
These expressions enable us to calculate the explicit components of the Weyl tensor for any non-twisting and shear-free geometry of an arbitrary dimension $D$. After a straightforward but very lengthy calculation we obtain
\begin{eqnarray}
&& \fl C_{rprq} = 0 \,, \label{Weyl rprq} \\
&& \fl C_{rpru} = {\textstyle \frac{D-3}{D-2}\big[-\frac{1}{2}g_{up,rr}
+g_{up}\Theta_{,r} +\Theta_{,p}+\Theta g_{up,r} \big]} \,, \\
&& \fl C_{rpmq} = {\textstyle \frac{2}{D-2}\big[-\frac{1}{2}g_{p[m}g_{q]u,rr}+g_{p[m}g_{q]u}\Theta_{,r}+g_{p[m}\Theta_{,q]}+\Theta g_{p[m}g_{q]u,r}\big]} \,,\\
&& \fl C_{ruru} = {\textstyle -\frac{D-3}{D-1}\Big[\big(\frac{1}{2}g_{uu,r}-\Theta g_{uu}\big)_{,r}+\frac{1}{(D-2)(D-3)}\,^{S}R} \nonumber \\
&& \fl \hspace{25.0mm} {\textstyle -\frac{1}{4}\frac{D-4}{D-2}g^{mn}g_{um,r}g_{un,r}+\frac{1}{D-2}\big(g^{rn}g_{un,rr}+g^{mn}g_{um,r||n}\big)} \nonumber \\
&& \fl \hspace{25.0mm} {\textstyle -\frac{2}{D-2}g^{rn}g_{un}\Theta_{,r}-2\Theta_{,u}-\frac{4}{D-2}g^{rn}\Theta_{,n}} \nonumber \\
&& \fl \hspace{25.0mm} {\textstyle -\Theta^2\frac{D-4}{D-2}g^{rn}g_{un}+\Theta\big(\frac{D-6}{D-2}g^{rn}g_{un,r}-\frac{2}{D-2}g^{mn}g_{um||n}\big)\Big]} \,, \\
&& \fl C_{rpuq} = {\textstyle \frac{1}{D-2}\Big[\,^{S}R_{pq}-\frac{1}{D-1}\,g_{pq}\,^{S}R+\frac{1}{2}(D-2)g_{u[p,r||q]}+\frac{1}{2}(D-4)f_{pq}} \nonumber \\
&& \fl \hspace{30.0mm} {\textstyle +\frac{1}{2}\frac{D-3}{D-1}g_{pq}g_{uu,rr}-\frac{1}{2}g_{up}g_{uq,rr}-\frac{1}{2}\frac{D-5}{D-1}g_{pq}g^{rn}g_{un,rr}} \nonumber \\
&& \fl \hspace{30.0mm} {\textstyle -\frac{1}{2}\frac{D-4}{D-1}g_{pq}g^{mn}g_{um,r}g_{un,r}-\frac{1}{2}\frac{D-5}{D-1}g_{pq}g^{mn}g_{um,r||n}} \nonumber \\
&& \fl \hspace{22.0mm} {\textstyle +\Theta_{,r}\big(g_{up}g_{uq}-\frac{D-3}{D-1}g_{pq}g_{uu}+\frac{D-5}{D-1}g_{pq}g^{rn}g_{un}\big)-2\frac{D-3}{D-1}g_{pq}\Theta_{,u}} \nonumber \\
&& \fl \hspace{22.0mm} {\textstyle +2\frac{D-5}{D-1}g_{pq}g^{rn}\Theta_{,n}-(D-5)g_{u(p}\Theta_{,q)}-(D-3)g_{u[p}\Theta_{,q]}} \nonumber \\
&& \fl \hspace{22.0mm} {\textstyle +\Theta^2(D-4)\big(g_{up}g_{uq}-\frac{2}{D-1}g_{pq}g^{rn}g_{un}\big)} \nonumber \\
&& \fl \hspace{22.0mm} {\textstyle +\Theta\big(\frac{3D-13}{D-1}g_{pq}g^{rn}g_{un,r}-\frac{D-3}{D-1}g_{pq}g_{uu,r}-(D-5)g_{u(p}g_{q)u,r}+g_{u[p}g_{q]u,r}} \nonumber \\
&& \fl \hspace{30.0mm} {\textstyle -(D-4)g_{u(p||q)}-(D-2)g_{u[p,q]}+\frac{D-5}{D-1}g_{pq}g^{mn}g_{um||n}\big)\Big]} \,, \\
&& \fl C_{rupq} = {\textstyle g_{u[p,q],r}-\frac{1}{D-2}g_{u[p}g_{q]u,rr}} \nonumber \\
&& \fl \hspace{15.0mm} {\textstyle -2\frac{D-3}{D-2}g_{u[p}\Theta_{,q]}-2\Theta\big(g_{u[p,q]}-\frac{1}{D-2}g_{u[p}g_{q]u,r}\big)} \,, \\
&& \fl C_{mpnq} = {\textstyle \,^{S}C_{mpnq}+\frac{2}{(D-2)(D-4)}\big(g_{mn}\,^{S}R_{pq}+g_{pq}\,^{S}R_{mn}-g_{mq}\,^{S}R_{pn}-g_{pn}\,^{S}R_{mq}\big)} \nonumber \\
&& \fl \hspace{27.5mm} {\textstyle +\frac{1}{D-2}\big(g_{mn}f_{pq}+g_{pq}f_{mn}-g_{mq}f_{pn}-g_{pn}f_{mq}\big)} \nonumber \\
&& \fl \hspace{15.0mm} {\textstyle -\frac{2}{D-2}\Big[\,g_{mn}\big(g_{u(p}\Theta_{,q)}-\Theta^2g_{up}g_{uq}+\Theta(g_{u(p||q)}+g_{u(p}g_{q)u,r}) \big)} \nonumber \\
&& \fl \hspace{25.5mm} {\textstyle +g_{pq}\big(g_{u(m}\Theta_{,n)}-\Theta^2g_{um}g_{un}+\Theta(g_{u(m||n)}+g_{u(m}g_{n)u,r}) \big)} \nonumber \\
&& \fl \hspace{24.5mm} {\textstyle -g_{mq}\big(g_{u(p}\Theta_{,n)}-\Theta^2g_{up}g_{un}+\Theta(g_{u(p||n)}+g_{u(p}g_{n)u,r}) \big)} \nonumber \\
&& \fl \hspace{25.5mm} {\textstyle -g_{pn}\big(g_{u(m}\Theta_{,q)}-\Theta^2g_{um}g_{uq}+\Theta(g_{u(m||q)}+g_{u(m}g_{q)u,r}) \big)\Big]} \nonumber \\
&& \fl \hspace{15.0mm} {\textstyle +\frac{1}{(D-1)(D-2)}\,(g_{mn}g_{pq}-g_{mq}g_{pn})\Big[\,g_{uu,rr}-2g^{rs}g_{us,rr}}\nonumber \\
&& \fl \hspace{25.0mm} {\textstyle
-2g^{os}f_{os}-\frac{1}{2}g^{os}g_{uo,r}g_{us,r}-\frac{2(2D-5)}{(D-3)(D-4)}\,^{S}R} \nonumber \\
&& \fl \hspace{25.0mm} {\textstyle +2\Theta_{,r}(-g_{uu}+2g^{rs}g_{us})-4\Theta_{,u}+8g^{rs}\Theta_{,s}} \nonumber \\
&& \fl \hspace{25.0mm} {\textstyle -6\Theta^2\,g^{rs}g_{us}+2\Theta\,\big(-g_{uu,r}+5g^{rs}g_{us,r}+2g^{os}g_{uo||s}\big)\Big]} \,,\\
&& \fl C_{ruup} = {\textstyle \frac{1}{2}\frac{D-3}{D-2}(g_{uu,rp}-g_{up,ru})+\frac{1}{4}\frac{D-4}{D-2}g^{rn}g_{un,r}g_{up,r}-\frac{1}{2}\frac{D-3}{D-2}g^{mn}g_{um,r}E_{np}} \nonumber \\
&& \fl \hspace{12.5mm} {\textstyle +\frac{1}{D-2}\big[g^{mn}e_{m[p}g_{n]u,r}+g^{mn}\big(g_{m[p,u||n]}+g_{u[m,p]||n}\big)\big]} \nonumber \\
&& \fl \hspace{12.5mm} {\textstyle -\frac{1}{D-2}\big[\frac{1}{2}g^{rn}g_{un}g_{up,rr}-g^{rn}g_{u[n,p],r}+\frac{1}{2}g^{rn}g_{up,r||n}\big]} \nonumber \\
&& \fl \hspace{12.5mm} {\textstyle -\frac{1}{(D-1)(D-2)}g_{up}\big[\,^{S}R-\frac{1}{2}(D-3)g_{uu,rr}+\frac{1}{2}(D-4)g^{mn}g_{um,r}g_{un,r}} \nonumber \\
&& \fl \hspace{39.5mm} {\textstyle +\frac{1}{2}(D-5)\big(g^{rn}g_{un,rr}+g^{mn}g_{um,r||n}\big)\big]} \nonumber \\
&& \fl \hspace{12.5mm} {\textstyle +\frac{D-3}{(D-1)(D-2)}\big[g_{up}\big(2g^{rn}g_{un}-g_{uu}\big)\Theta_{,r}+(D-3)g_{up}\Theta_{,u}} \nonumber \\
&& \fl \hspace{33.5mm} {\textstyle -(D-1)g_{uu}\Theta_{,p}+4g_{up}g^{rn}\Theta_{,n}\big]} \nonumber \\
&& \fl \hspace{12.5mm} {\textstyle +\Theta^{2}\frac{(D-3)(D-4)}{(D-1)(D-2)}g^{rn}g_{un}g_{up}} \nonumber \\
&& \fl \hspace{12.5mm} {\textstyle -\frac{1}{D-2}\Theta\big[(D-3)(g_{uu,p}-g_{up,u})+\frac{1}{2}(D-6)g^{rn}g_{un}g_{up,r}} \nonumber \\
&& \fl \hspace{26.5mm} {\textstyle -\frac{1}{2}(D-2)g^{rn}g_{np,u}-(D-6)g^{rn}g_{u[n,p]}\big]} \nonumber \\
&& \fl \hspace{12.5mm} {\textstyle -\frac{1}{(D-1)(D-2)}g_{up}\Theta\big[\frac{1}{2}(D-5)(D-6)g^{rn}g_{un,r}+\frac{1}{2}(D-1)g^{mn}g_{mn,u}} \nonumber \\
&& \fl \hspace{39.5mm} {\textstyle +(D-3)\big(g_{uu,r}-2g^{mn}g_{um||n}\big)\big]} \,, \\
&& \fl C_{upmq} = {\textstyle g_{p[m,u||q]}+g_{u[q,m]||p}+e_{p[m}g_{q]u,r}+\frac{2}{D-2}\big(\,^{S}R_{p[m}g_{q]u}-f_{p[m}g_{q]u}\big)} \nonumber \\
&& \fl \hspace{12.5mm} {\textstyle +\frac{1}{D-2}\Big[(g_{uu}-g^{rn}g_{un})g_{p[m}g_{q]u,rr}-g_{uu,r[q}g_{m]p}+g_{p[m}g_{q]u,ru}} \nonumber \\
&& \fl \hspace{24mm} {\textstyle +g^{rn}(g_{pm}g_{u[n,q],r}-g_{pq}g_{u[n,m],r})-g^{rn}g_{un,r}g_{p[m}g_{q]u,r}-g^{rn}g_{p[m}g_{q]u,r||n}} \nonumber \\
&& \fl \hspace{24mm} {\textstyle +g^{ns}g_{un,r}g_{us||[q}g_{m]p}+g^{ns}(g_{pm}g_{n[q,u||s]}-g_{pq}g_{n[m,u||s]})}
\nonumber \\
&& \fl \hspace{24mm} {\textstyle +g^{ns}(g_{pm}g_{u[n,q]||s}-g_{pq}g_{u[n,m]||s})-g^{ns}e_{ns}g_{p[m}g_{q]u,r}\Big]} \nonumber \\
&& \fl \hspace{12.5mm} {\textstyle -\frac{2}{(D-1)(D-2)}\,g_{p[m}g_{q]u}\big[\,^{S}R+g_{uu,rr}-2g^{rn}g_{un,rr}-\frac{3}{2}g^{ns}g_{un,r}g_{us,r}} \nonumber \\
&& \fl \hspace{24mm} {\textstyle -2g^{ns}g_{un,r||s}
+\big((D-3)g_{uu}-(D-5)g^{rn}g_{un}\big)\Theta_{,r}} \nonumber\\
&& \fl \hspace{24mm} {\textstyle +(D-5)\Theta_{,u}-2(D-5)g^{rn}\Theta_{,n}\big]} \nonumber \\
&& \fl \hspace{12.5mm} {\textstyle +\frac{2}{D-2}g_{up}\Theta_{,[m}g_{q]u}-4\Theta^2\frac{D-4}{(D-1)(D-2)}g^{rn}g_{un}g_{p[m}g_{q]u}-2\Theta e_{p[m}g_{q]u}} \nonumber \\
&& \fl \hspace{12.5mm} {\textstyle +\frac{2}{D-2}\Theta\big[g_{u(p||m)}g_{uq}-g_{u(p||q)}g_{um}-g_{up}g_{u[m}g_{q]u,r}-g_{p[m}g_{q]u,r}(g_{uu}-2g^{rn}g_{un})} \nonumber \\
&& \fl \hspace{24mm} {\textstyle -2g^{rn}(g_{pm}g_{u[n,q]}-g_{pq}g_{u[n,m]})-g_{p[m}g_{q]u,u}+g_{uu,[q}g_{m]p}\big]} \nonumber \\
&& \fl \hspace{12.5mm} {\textstyle +\frac{2}{(D-1)(D-2)}\Theta\, g_{p[m}g_{q]u} \big[2g_{uu,r}+(D-11)g^{rn}g_{un,r}} \nonumber \\
&& \fl \hspace{48.6mm} {\textstyle +(D-5)g^{ns}g_{un||s}-\frac{1}{2}(D-1)g^{ns}g_{ns,u}\big]} \,,\\
&& \fl C_{upuq} = {\textstyle -\frac{1}{2}g_{uu||p||q}-\frac{1}{2}g_{pq,uu}+g_{u(p,u||q)}-\frac{1}{2}g_{uu,r}e_{pq}+\frac{1}{2}g_{uu,(p}g_{q)u,r}+g^{os}E_{op}E_{sq}} \nonumber \\
&& \fl \hspace{12.5mm} {\textstyle -\frac{1}{D-2}\,g_{pq}g^{mn}\big(-\frac{1}{2}g_{uu||m||n}-\frac{1}{2}g_{mn,uu}+g_{um,u||n}} \nonumber \\
&& \fl \hspace{30.5mm} {\textstyle -\frac{1}{2}g_{uu,r}e_{mn}+\frac{1}{2}g_{uu,m}g_{un,r}+g^{os}E_{om}E_{sn}\big)} \nonumber \\
&& \fl \hspace{12.5mm} {\textstyle +\frac{1}{(D-1)(D-2)}\,(g_{uu}g_{pq}-g_{up}g_{uq})\big(\,^{S}R+g_{uu,rr}-2g^{rn}g_{un,rr}} \nonumber\\
&& \fl \hspace{30.5mm} {\textstyle
-\frac{3}{2}g^{mn}g_{um,r}g_{un,r}-2g^{mn}g_{um,r||n}\big)} \nonumber \\
&& \fl \hspace{12.5mm} {\textstyle -\frac{1}{2(D-2)}\,g_{uu}g_{pq}\big(g_{uu,rr}-g^{mn}g_{um,r}g_{un,r}\big)
-\frac{1}{D-2}\,g_{uu}\,^{S}R_{pq}} \nonumber \\
&& \fl \hspace{12.5mm} {\textstyle -\frac{1}{4}\big(\frac{D-4}{D-2}g_{uu}-g^{rn}g_{un}\big)g_{up,r}g_{uq,r}
+\frac{1}{D-2}\,g_{uu}g_{u(p,r||q)}-g^{rn}E_{n(p}g_{q)u,r}} \nonumber \\
&& \fl \hspace{12.5mm} {\textstyle +\frac{1}{D-2}\,g_{pq}g^{rn}\big[\frac{1}{2}g_{un}g_{uu,rr}+g_{uu,rn}-g_{un,ru}} \nonumber \\
&& \fl \hspace{31.0mm} {\textstyle -\frac{1}{2}g_{uo,r}\big(g^{os}g_{us,r}g_{un}-g^{ro}g_{un,r}-4g^{os}g_{u[n,s]}\big)\big]} \nonumber \\
&& \fl \hspace{12.5mm} {\textstyle +\frac{1}{D-2}\big[(g_{uu}-g^{rn}g_{un})g_{u(q}g_{p)u,rr}-g_{uu,r(p}g_{q)u}+g_{u(q}g_{p)u,ru}} \nonumber \\
&& \fl \hspace{25.5mm} {\textstyle +g^{rn}g_{un,r||(p}g_{q)u}-2g^{rn}g_{u(q}g_{p)u,r||n}-g^{rn}g_{un,r}g_{u(q}g_{p)u,r}\big]} \nonumber \\
&& \fl \hspace{12.5mm} {\textstyle +\frac{1}{D-2}\,g^{mn}\big[g_{um,r}g_{un||(p}g_{q)u}-e_{mn}g_{u(q}g_{p)u,r}+g_{u(q}g_{p)m,u||n}-g_{mn,u||(p}g_{q)u}} \nonumber \\
&& \fl \hspace{30.5mm} {\textstyle +\frac{1}{2}\big(g_{uq}g_{um||p||n}+g_{up}g_{um||q||n}\big)-g_{u(q}g_{p)u||m||n}\big]} \nonumber \\
&& \fl \hspace{12.5mm} {\textstyle +\frac{1}{(D-1)(D-2)}\Theta_{,r}\big[ g_{uu}g_{pq}\big((D-3)g_{uu}-(D-5)g^{rn}g_{un}\big)} \nonumber \\
&& \fl \hspace{34.7mm} {\textstyle -2g_{up}g_{uq}
\big((D-2)g_{uu}-(D-3)g^{rn}g_{un}\big)\big]} \nonumber \\
&& \fl \hspace{12.5mm} {\textstyle +\Theta_{,u}\frac{2(D-3)}{(D-1)(D-2)}\big(g_{uu}g_{pq}-g_{up}g_{uq}\big)-\frac{2}{D-2}\,g_{uu}g_{u(p}\Theta_{,q)}} \nonumber \\
&& \fl \hspace{12.5mm} {\textstyle -2g^{rn}\Theta_{,n}\big(\frac{D-5}{(D-1)(D-2)}g_{uu}g_{pq}-\frac{2(D-3)}{(D-1)(D-2)}g_{up}g_{uq}\big)} \nonumber \\
&& \fl \hspace{12.5mm} {\textstyle +\Theta^2\big[\frac{2(D-4)}{(D-1)(D-2)}g^{rn}g_{un}g_{uu}g_{pq}-\frac{D-4}{D-2}g_{uu}g_{up}g_{uq}+\frac{(D-3)(D-4)}{(D-1)(D-2)}g^{rn}g_{un}g_{up}g_{uq}\big]} \nonumber \\
&& \fl \hspace{12.5mm} {\textstyle +\Theta\big[\frac{1}{D-2}\,g_{pq}g^{rn}\big(2g_{un,u}-g_{uu,n}-g_{un}g_{uu,r}\big)+\frac{D-6}{D-2}\big(g_{uu}-g^{rn}g_{un}\big)g_{u(q}g_{p)u,r}} \nonumber \\
&& \fl \hspace{18.5mm} {\textstyle +\frac{D-3}{(D-1)(D-2)}g_{uu,r}g_{uu}g_{pq}+\frac{2}{(D-1)(D-2)}\,g_{up}g_{uq}g_{uu,r}} \nonumber \\
&& \fl \hspace{18.5mm} {\textstyle -\frac{2(D-6)}{(D-1)(D-2)}\,g^{rn}g_{un,r}\big(g_{uu}g_{pq}-g_{up}g_{uq}\big)} \nonumber \\
&& \fl \hspace{18.5mm} {\textstyle +g_{uu}e_{pq}+2g^{rn}E_{n(p}g_{q)u}-\frac{1}{D-2}\,g_{uu}\big(2g_{u(p||q)}-g_{pq}g^{mn}e_{mn}\big)} \label{Weyl upuq} \\
&& \fl \hspace{18.5mm} {\textstyle -\frac{2}{D-2}g_{u(q}g_{p)u,u}-\frac{D-4}{D-2}g_{uu,(p}g_{q)u}-\frac{4}{D-2}g^{rn}\big(g_{un||(p}g_{q)u}-g_{u(q}g_{p)u||n}\big)} \nonumber \\
&& \fl \hspace{18.5mm} {\textstyle +\frac{1}{(D-1)(D-2)}(g_{uu}g_{pq}-g_{up}g_{uq})g^{mn}\big((D-1)g_{mn,u}-2(D-3)g_{um||n}\big)\big]} \,. \nonumber
\end{eqnarray}

In the above expressions, ${\,^{S}\Gamma^m_{pq}\equiv\frac{1}{2}g^{mn}(2g_{n(p,q)}-g_{pq,n})}$ denote Christoffel symbols with respect to the spatial coordinates only, i.e., the coefficients of the covariant derivative on the transverse ${(D-2)}$-dimensional Riemannian space. The symbol ${\,_{||}}$ denotes this covariant derivative with respect to $g_{pq}$.
Similarly, ${\,^{S}R_{mpnq}}$, ${\,^{S}C_{mpnq}}$, ${\,^{S}R_{pq}}$ and ${\,^{S}R}$ are the Riemann tensor, Weyl tensor, Ricci tensor and Ricci scalar for the transverse-space metric ${g_{pq}}$, respectively. We have also introduced the following useful auxiliary quantities\,:
\begin{eqnarray}
g_{up||q} \erovno g_{up,q}-g_{um}\,^{S}\Gamma^{m}_{pq} \,, \label{auxiliaryB}\\
g_{up,r||q} \erovno g_{up,rq}-g_{um,r}\,^{S}\Gamma^{m}_{pq} \,, \\
g_{u[p,r||q]} \rovno g_{u[p||q],r}\,, \\
g_{p[m,u||q]} \erovno g_{p[m,q],u}+{\textstyle \frac{1}{2}}(\,^{S}\Gamma^{n}_{pm}\,g_{nq,u}-\,^{S}\Gamma^{n}_{pq}\,g_{nm,u}) \,,\\
g_{u[q,m]||p} \erovno g_{u[q,m],p}-\,^{S}\Gamma^{n}_{pq}\,g_{u[n,m]}-\,^{S}\Gamma^{n}_{pm}\,g_{u[q,n]} \,,\\
(g_{uu})_{||p||q} \erovno g_{uu,pq}-g_{uu,n}\,^{S}\Gamma^{n}_{pq}\,, \\
g_{up,u||q} \erovno g_{up,uq}-g_{um,u}\,^{S}\Gamma^{m}_{pq} \,,\\
e_{pq} \erovno g_{u{(p||q)}}- {\textstyle \frac{1}{2}}g_{pq,u} \,, \\
E_{pq} \erovno g_{u{[p,q]}}+ {\textstyle \frac{1}{2}}g_{pq,u} \,, \\
f_{pq} \erovno g_{u(p,r||q)}+ {\textstyle \frac{1}{2}}g_{up,r}g_{uq,r} \,,
\label{auxiliaryE}
\end{eqnarray}
where ${g_{u{[p,q]}}=g_{u{[p||q]}}}$. These are tensors on the transverse Riemannian space.


\section*{References}

\begin{thebibliography}{10}

\bibitem{RobTra60}
Robinson I and Trautman A 1960 Spherical gravitational waves
{\em Phys. Rev. Lett.} {\bf 4} 431--2

\bibitem{RobTra62}
Robinson I and Trautman A 1962 Some spherical gravitational waves in general relativity
{\em Proc. Roy. Soc.~A} {\bf 265} 463--73

\bibitem{Kundt:1961}
Kundt W 1961 The plane-fronted gravitational waves
{\em Z.~Physik} {\bf 163} 77--86

\bibitem{Kundt:1962}
Kundt W 1962 Exact solutions of the field equations: twist-free pure radiation
fields {\em Proc. Roy. Soc.~A} {\bf 270} 328--34

\bibitem{Stephani:2003}
Stephani H, Kramer D, MacCallum M, Hoenselaers C and Herlt E 2003 {\em Exact Solutions of Einstein's Field Equations}
(Cambridge: Cambridge University Press)

\bibitem{GriffithsPodolsky:2009}
Griffiths J B and Podolsk\'{y} J 2009 {\em Exact Space-Times in Einstein's General Relativity}
(Cambridge: Cambridge University Press)

\bibitem{PodOrt06}
Podolsk\'y J and Ortaggio M 2006 Robinson--Trautman spacetimes in higher dimensions
{\em Class. Quantum Grav.} {\bf 23} 5785--97

\bibitem{OrtPodZof08}
Ortaggio M, Podolsk\'y J and \v{Z}ofka M 2008 Robinson--Trautman spacetimes with an electromagnetic field in higher dimensions
{\em Class. Quantum Grav.} {\bf 25} 025006 (18pp)

\bibitem{SvarcPodolsky:2014}
\v{S}varc~R and Podolsk\'{y}~J 2014 Absence of gyratons in the Robinson--Trautman class,
{\em Phys. Rev.~D}, in press (arXiv 1406.0729).

\bibitem{PodolskyZofka:2009}
Podolsk\'{y} J and \v{Z}ofka M 2009 General Kundt spacetimes in higher dimensions
{\em Class. Quantum Grav.} {\bf 26} 105008 (18pp)

\bibitem{ColeyEtal:2009}
Coley~A, Hervik~S, Papadopoulos~G and Pelavas~N 2009 Kundt spacetimes
{\em Class. Quantum Grav.} {\bf 26} 105016 (34pp)

\bibitem{OrtaggioPravdaPravdova:2013}
Ortaggio M, Pravda V and Pravdov\'a A 2013 Algebraic classification of higher dimensional spacetimes based on null alignment
{\em Class. Quantum Grav.} {\bf 30} 013001 (57pp)

\bibitem{PodolskySvarc:2013a}
Podolsk\'{y}~J and \v{S}varc~R 2013 Explicit algebraic classification of Kundt geometries in any dimension
{\em Class. Quantum Grav.} {\bf 30} 125007 (25pp)

\bibitem{PodolskySvarc:2013b}
Podolsk\'{y}~J and \v{S}varc~R 2013 Physical interpretation of Kundt spacetimes using geodesic deviation
{\em Class. Quantum Grav.} {\bf 30} 205016 (24pp)

\bibitem{PodolskySvarc:2012}
Podolsk\'{y}~J and \v{S}varc~R 2012 Interpreting spacetimes of any dimension
using geodesic deviation
{\em Phys. Rev.~D} {\bf 85} 044057 (18pp)

\bibitem{ColeyMilsonPravdaPravdova:2004}
Coley~A, Milson R, Pravda V and Pravdov\'a A 2004 Classification of the Weyl tensor in higher dimensions {\em Class. Quantum Grav.} {\bf 21} L35--41

\bibitem{Wald:1984}
Wald~R~M 1984 {\em General Relativity}
(Chicago: University of Chicago Press)

\end{thebibliography}
\end{document}